\documentclass[10pt,onecolumn,draft,final]{IEEEtran}

\usepackage{amsmath,amssymb}
\usepackage{danmacros}
\usepackage{times}
\usepackage{mathbf-abbrevs}
\usepackage{citesort}
\usepackage{graphicx}
\usepackage{units}
\usepackage{fancyhdr}
\usepackage{amsbsy}

\renewcommand{\baselinestretch}{1.75}

\renewcommand{\xML}{\xhbf^{\text{opt}}}
\newcommand{\xho}{\xh^{\text{opt}}}
\newcommand{\xhub}{\underline{\xh}}
\newcommand{\xhubo}{\xhub^{\text{opt}}}
\newcommand{\xbest}{\xhbf^{\text{best}}}%
\newcommand{\xubML}{\underline{\xhbf}^{\text{opt}}}%
\newcommand{\xhubf}{\underline{\xhbf}}%
\renewcommand{\hML}{\smat{h}^{\text{opt}}}
\newcommand{\lambdaML}{\smat{\lambda}^{\text{opt}}}
\newcommand{\lambdabfML}{\smat{\lambdabf}^{\text{opt}}}

\newcommand{\unbf}[1]{\underline{\textbf{#1}}}
\newcommand{\Lmax}{L^{\text{max}}}

\newcommand{\Bmax}{B^{\text{max}}}
\newcommand{\lambdamax}{\lambda^{\text{max}}}
\newcommand{\ca}{\addtocounter{a}{1}\arabic{a}}
\newcommand{\algnum}{\scriptsize{\bf{\ca}}&\scriptsize}
\newcommand{\algnon}{&\scriptsize}

\begin{document}
\newcounter{a}

\vspace{1cm}
\title{GLRT-Optimal Noncoherent Lattice Decoding}

\vspace{2cm}
\author{
\authorblockN{{\bf Daniel~J.~Ryan$^{\dag \ddag *}$, ~Iain~B.~Collings$^\ddag$~ and ~I.~Vaughan~L.~Clarkson$^\star$~
\thanks{
I.~V.~L.~Clarkson was on study leave throughout 2005 at the Dept. of Electrical \& Computer Engineering, University of British Columbia, Canada. This paper has appeared in part at VTC-Spring 2006 and ICASSP 2006. }
}
}\\
\small
\vspace{1cm}
\footnotesize \authorblockA{$^\dag$School of Electrical and Information Engineering, University of Sydney, \textsc{Australia}, \\
Phone: +612 9372 4465, Fax: +612 9372 4490, Email: \texttt{dan@ee.usyd.edu.au} \\
}
\authorblockA{$^\ddag$Wireless Technologies Laboratory, CSIRO ICT Centre, \textsc{Australia},\\
Phone: +612 9372 4120, Fax: +612 9372 4490, Email: \texttt{iain.collings@csiro.au}\\
}
\authorblockA{$^\star$School of Information Technology and Electrical Engineering, University of Queensland, \textsc{Australia}, \\
Phone: +617 3365 8834, Fax: +617 3365 4999, Email: \texttt{v.clarkson@itee.uq.edu.au}\\
}
}

\vspace{-1.8cm}

%
\markboth{IEEE Transactions on Signal Processing, accepted to appear (accepted Nov. 2006)}{\papertitle}
%
\maketitle

\begin{abstract}
This paper presents new low-complexity lattice-decoding algorithms for \emph{noncoherent} block detection of QAM and PAM signals over complex-valued fading channels. The algorithms are optimal in terms of the generalized likelihood ratio test (GLRT). The computational complexity is \emph{polynomial} in the block length; making GLRT-optimal noncoherent detection feasible for implementation.  We also provide even lower complexity suboptimal algorithms.  Simulations show that the suboptimal algorithms have performance indistinguishable from the optimal algorithms. Finally, we consider block based transmission, and propose to use noncoherent detection as an alternative to pilot assisted transmission (PAT). The new technique is shown to outperform PAT.
\end{abstract}

\section{Introduction} \label{sec:Intro}
\normalsize
Noncoherent transmission of digital signals over unknown fading channels has recently received significant attention especially for the case of the block-fading channel model. Applications include recovery from deep fades in pilot-symbol assisted modulation based schemes, eavesdropping, and non-data-aided channel estimation. Noncoherent transmission is particularly applicable to systems exhibiting small coherence intervals where the use of training signals would result in a significant loss in throughput. Recently, some elegant information-theoretic results have been derived for noncoherent single and multiple-antenna systems under the assumption of Rayleigh fading, for example \cite{Marzetta99,Zheng02}. Information theoretic aspects of noncoherent transmission were considered in \cite{Hassibi03} which concluded that at low SNR and for small coherence intervals there is a significant capacity penalty by using training. Under this noncoherent detection regime, it has been shown by numerical simulation that standard modulation techniques such as quadrature amplitude modulation (QAM) can achieve near-capacity in the single-antenna noncoherent block Rayleigh-fading channel \cite{Chen03}.

This paper focuses on noncoherent receiver design for the block fading channel. A great deal of work has been performed on partially coherent receivers such as pilot-symbol assisted modulation (PSAM) \cite{Cavers91,Tong04}, per-survivor techniques \cite{Raheli95}, and coupled estimators \cite{Davis01b}. However the challenge remains to develop high-performance, low-complexity, fully noncoherent receivers.

Various suboptimal algorithms have been proposed for block-based noncoherent detection. For slowly fading channels, a blind phase recovery approach was proposed in \cite{Georghiades97} for noncoherent detection of differentially encoded QAM \cite{Weber78} where the attenuation was assumed to be known exactly at the receiver. In \cite{Warrier02}, a suboptimal technique for PSK was proposed which involved forming a number of equally spaced channel phase estimates. An extension to multi-amplitude constellations was also presented, where every sequence of symbol amplitudes is considered, and then the PSK technique is applied to determine the phase of the symbols. Unfortunately, the complexity of this suboptimal approach is still exponential in the sequence length, albeit with a smaller base.

Recently, lattice decoding algorithms have been applied to noncoherent and differential detection. For PSK over temporally-correlated Rayleigh fading channels, a form of lattice decoding (namely sphere-decoding) can be applied since it turns out that the detection metric is Euclidean \cite{Lampe05}.  Lattice decoding techniques have also been used for differential detection of diagonal space-time block codes over Rayleigh fading channels, by approximating the decision metric with a Euclidean metric \cite{ClarksonK01,CongLing05}. In \cite{RyanISIT05}, we presented simulation results for another lattice projection approach for suboptimal PAM and QAM detection. Unfortunately, each of these algorithms require complexity exponential with the block length to guarantee that the optimal estimate is found \cite{PauliISIT06,Agrell2002}.  Practical implementation considerations demand that low complexity algorithms be developed.

\pubidadjcol

For the case of the constant envelope PSK constellation, a detection algorithm with complexity $O(T \log T)$ was developed in \cite{Mackenthun,Sweldens01} (where $T$ is the block length), which can provide the optimal data estimate over an unknown noncoherent fading channel, in terms of the Generalized Likelihood Ratio Test (GLRT). The GLRT is equivalent to joint ML estimation of a continuous valued channel parameter and discrete-valued data parameters. This approach was generalized in \cite{Motedayen03}, where they outlined a general graph-based approach which involved forming a spanning tree.  Specific details were presented for the cases of QAM over a phase noncoherent channel (\ie known channel amplitude) \cite{Motedayen03,MotedayenICC02}, and for PSK over a fading channel with coding \cite{MotedayenICC03}.  The challenge remains to develop efficient algorithms for optimal noncoherent sequence detection of multi-amplitude constellations over fading channels.

In this paper we propose a new GLRT-optimal noncoherent lattice decoding approach for QAM and PAM symbols which has complexity polynomial in the block length. We start by considering detection of $M$-ary PAM over real-valued fading channels (which we shall term real-PAM). We show that the GLRT-optimal codeword estimate is the closest codeword (or lattice point) \emph{in angle} to the \emph{line} described by the received vector. We propose an algorithm that searches along the line, and chooses the best codeword estimate from this search. We provide a theorem that bounds the search to a segment of the line, limiting the number of codewords that need to be considered.  We show how the search can be done in an iterative manner, and that the resulting complexity of the algorithm is $O(T \log T)$.

We then consider the more practical case of $M$-ary QAM detection over complex-valued fading channels, and show that in this case the GLRT-optimal codeword estimate is the closest codeword in angle to a \emph{plane} described by the received vector. We propose an algorithm that searches across the plane, and chooses the best codeword estimate from this search. We provide a theorem that bounds the search to a segment of the plane. We show how the extent of the search can be further reduced by exploiting the rotational symmetry of the constellation. The resulting plane search algorithm can be performed with complexity of order $O(T^3)$.

We also present new suboptimal noncoherent QAM detection algorithms with even lower complexity; by combining a channel phase estimator with our fast real-PAM algorithm. We propose using $O(T)$ instances of the real-PAM algorithm.  This approach therefore has complexity of order $O(T^2 \log T)$. Simulations indicate that there is a negligible performance loss compared to GLRT, when using this suboptimal technique.

Finally, we also propose a pilot-assisted version of our new reduced-search noncoherent lattice-decoding algorithms. The pilot symbol is used to remove the ambiguities inherent with noncoherent detection. Our approach obtains improved performance compared with standard pilot assisted transmission \cite{Tong04}, while maintaining the same data rate.

\section{System Model}
\subsection{Signal Model} \label{sec:SigModel}

We define a codebook $\Ccal(\Xcal,T)$ as the set of all possible sequences of $T$ transmitted symbols, $\xbf = [ x_1,\ldots,x_T ]'$, such that each $x_t$ is in some constellation $\Xcal$. For an $M$-ary PAM constellation, $\Xcal = \cubr{\pm 1, \pm 3,\ldots, \pm(M-1)}$. For QAM, $\Xcal$ is a subset of the Gaussian (complex) integers with odd real and imaginary components. For example, for an $M^2$-ary square QAM constellation, $\Xcal = \cubr{ \; x \; | \Real{x} \in \Xcal', \; \Imag{x} \in \Xcal'}$ where $\Xcal' \defas \cubr{\pm 1, \pm 3,\ldots, \pm (M-1)}$. Thus each codebook $\Ccal(\Xcal,T)$ is a set of lattice points drawn from a subset of the unit lattice of $\Rbb^T$ or $\Cbb^T$.

We consider block fading channels and assume that the channel $h$ is constant for at least $T$ symbols as in \cite{Chen03,Zheng02,Hassibi03,Marzetta99,Warrier02}.  We will consider narrowband fading channels where $h$ is either real-valued and complex-valued channels. Thus we can write the received codeword $\ybf = [\; y_1,\ldots,y_T \;]'$ as follows,
\begin{equation}\label{eq:VecModel}
\ybf = h\xbf + \nbf
\end{equation}
where $\nbf = [ \; n_1,\ldots,n_T \; ]'$ is a vector of additive white Gaussian noise.

\subsection{Detection} \label{sec:Detection}
The noncoherent detection problem is to estimate $\xbf$ based on $\ybf$ without knowledge of the channel and in the absence of training data. The log-likelihood function of the maximum likelihood (ML) detector (of both channel and data) is given by
\begin{equation} \label{eq:LhoodEuc}
  \fn{L}{\ybf; \xbf, h }
    = -\magn{\ybf - h \xbf}^2
\end{equation}
where constant factors have been discarded and $\magn{\cdot}$ represents the Euclidean norm. For a given codeword hypothesis $\xhbf$, the likelihood function is maximized by choosing
\begin{equation}\label{eq:hest}
  \smat{h}
    = \frac{\xhbf^\dag \ybf}%
        {\magn{\xhbf}^2}
\end{equation}
where $(\cdot)^\dag$ denotes Hermitian transpose.

Hence, the ML estimate of $\xbf$ conditioned on the corresponding channel estimate, is given by
\begin{align}
\xML \nonumber
&=\arg \max_{\xhbf \in \Ccal(\Xcal,T)}
L \left ( \ybf; \xhbf, \frac{\xhbf^\dag \ybf} {\magn{\xhbf}^2} \right ) \\
&= \arg \max_{\xhbf \in \Ccal(\Xcal,T)}
\frac{ \abs{ \xhbf^\dag \ybf}^2}{ \magn{\xhbf}^2 }
\label{eq:x_ML}
\end{align}
This is the Generalized Likelihood Ratio Test (GLRT) \cite{VanTrees} considered in \cite{Warrier02,Motedayen03}.

Note that (\ref{eq:x_ML}) is equivalently given by
\begin{align}
\xML
&= \arg \max_{\xhbf \in \Ccal(\Xcal,T)} \frac{ \abs{ \xhbf ^\dag \ybf}^2} { \magn{\xhbf}^2 \magn{\ybf}^2} \\
&= \arg \max_{\xhbf \in \Ccal(\Xcal,T)} \cos^2 \theta(\xhbf,\ybf) \label{eq:cos_def}
\end{align}
where $\theta(\xbf,\ybf)$ is the principal angle between $\xbf$ and $\ybf$ \cite{Wong67}. Thus $\xML$ can be found by searching the points of $\Ccal(\Xcal,T)$ to find the one closest in angle to $\ybf$.

For QAM, we can also obtain a geometric interpretation of (\ref{eq:x_ML}) by expressing the complex vectors in $\Rbb^{2T}$.  We will use the underscore notation $\xubbf$ to denote the mapped version of $\xbf$ as follows,
\begin{equation} \label{eq:c2r}
\xubbf = [ \; \Real{x_1} \; \Imag{x_1} \; \ldots \; \Real{x_T} \; \Imag{x_T} ]'
\end{equation}
and denote the real-valued codebook as $\Ccal_R(\Xcal,T) = \cubr{ \xubbf \; | \; \xbf \in \Ccal }$.  For $M^2$-ary square QAM, we therefore have $\Ccal_R(\Xcal,T) = \Ccal(\Xcal',2T)$ where $\Xcal'$ is an $M$-ary PAM constellation. We also define $\Yubbf \in \Rbb^{2T \times 2}$ as a basis for the subspace $\ybf\Cbb$ mapped into the real space $\Rbb^{2T}$; that is
\begin{align} \label{eq:Ydef}
\Yubbf &\defas
\sqbr{
\begin{matrix}
\;\;\; \Real{y_1}   &\Imag{y_1} &\ldots & \;\;\;\Real{y_T} &\Imag{y_T} \\
-\Imag{y_1}   &\Real{y_1} &\ldots & -\Imag{y_T} &\Real{y_T}
\end{matrix}
}'.
\end{align}
Note that the columns of $\Yubbf$ are orthogonal. The projection matrix $\Pbf(\ybf) \in \Rbb^{2T \times 2T}$ is defined as
\begin{equation} \label{eq:Proj}
\Pbf(\ybf) \defas
\frac { \Yubbf \, \Yubbf' } { \magn{\ybf}^2 }
\end{equation}
such that
\begin{equation*}
\Pbf(\ybf) \xubbf  = \arg \min_{\vubbf \in \Yubbf\Rbb^2} \magn{\xubbf-\vubbf}.
\end{equation*}
That is, the vector $\Pbf(\ybf)\xubbf$ is the projection of $\xubbf$ onto the subspace $\Yubbf\Rbb^2$.
Now, it can be easily shown that
\begin{equation*}
\xubML
= \arg \max_{\hat{\xubbf}: \hat{\xbf}\in \Ccal(\Xcal,T)}
    \cos^2 \theta(\hat{\xubbf},\Pbf(\ybf)\hat{\xubbf})
\end{equation*}
Thus the GLRT-optimal data estimate $\xML$, corresponds to the $\hat{\xubbf} \in \Ccal_R(\Xcal,T)$ closest in angle to the plane $\Yubbf\Rbb^2$.

It is important to note that two forms of ambiguity exist for this noncoherent detection problem. The first is the well-known phase ambiguity which occurs for any constellation that is invariant to a particular phase rotation. For example, for square QAM constellations the following four optimal channel estimate and codeword pairs have the same likelihood: $(\hML, \xML)$, $(-\hML, -\xML)$, $(-i\hML, i\xML)$ and $(i\hML, -i\xML)$; corresponding to the four $\pi/2$ rotations of the constellation. We will assume that this type of ambiguity can be resolved, for example, by using the phase of the last symbol from the previous codeword \cite{Chen03}, or by using differential encoding \cite{Weber78}. The second type of ambiguity we call a divisor ambiguity and arises when there are multiple points in $\Ccal(\Xcal,T)$ that lie on the same 1-dimensional (real or complex) subspace e.g. $[1,1,1]$ and $[3,3,3]$ for 4-ary real-PAM with $T=3$. This produces a lower bound on the noncoherent block detection error rate as discussed and analyzed in \cite{RyanIT06}.

\section{Reduced Search Space}

In this section we show that the GLRT-optimal data estimate $\xML$, can be found without testing all the elements of $\Ccal(\Xcal,T)$. In the previous section we established that $\xML$ is the codeword closest in angle to a particular subspace, so it naturally makes sense to define a `nearest neighbor set' of the subspace and search within that set. The subspace of interest has basis vector $\ybf$ and passes through the origin. We show that the nearest neighbor set for this subspace contains $\xML$.  This implies that low complexity decoding algorithms can be developed, based on finding this particular nearest neighbor set, and searching it.

\begin{defn} \label{def:NN}
We define $NN(\vbf)$ to be the point, or set of points, in $\Ccal(\Xcal,T)$ closest to the arbitrary point $\vbf$ (\emph{i.e.}, the nearest neighbor to $\vbf$). That is, $\dbf$ is an element of $NN(\vbf)$ if $\magn{ \vbf - \dbf } \leq \magn{ \vbf - \zbf }$ for all $\zbf \in \Ccal(\Xcal,T)$.
\end{defn}
Of course, usually $NN(\vbf)$ will have a single element, and in this case we can write $NN(\vbf)=\dbf$.

\begin{defn} \label{def:Ncal}
We define $\Ncal(\Ccal(\Xcal,T),\ybf)$ to be the nearest neighbor subset of the codebook $\Ccal(\Xcal,T)$, corresponding to the subspace with basis vector $\ybf$, passing through the origin.  That is, $\ubf \in \fn{\Ncal}{\Ccal(\Xcal,T), \ybf}$ if and only if there exists some $\lambda$ such that $NN(\lambda \ybf)=\ubf$.
\end{defn}
Note that from a geometrical perspective, it is useful to think of $\lambda$ as being equivalent to the inverse of a channel estimate; implying that a point $\ubf$ is in the nearest neighbor set if there is a channel estimate $\hat{h}$ such that the distance  $|\ybf - \hat{h}\ubf|$ is smaller than for any other point.  Consequently we define $\lambdaML \defas (\hML)^{-1}$ as the reciprocal of the optimal channel estimate.

The following property of the GLRT-optimal codeword estimate $\xML$, allows us to reduce the set of codewords which need to be tested, to a small subset of the $\abs{\Xcal}^T$ possible codewords (where $\abs{\cdot}$ denotes set cardinality).  Note that an equivalent result was presented in \cite{Motedayen03}, however the geometrical interpretation of our formulation is more apparent; and is important when developing our new search algorithms later.
\begin{property}\label{prop:xML}
\begin{equation*}
\xML \in \Ncal(\Ccal(\Xcal,T),\ybf).
\end{equation*}
\end{property}
\begin{proof}
Consider the case where $\xML \notin \Ncal(\Ccal(\Xcal,T),\ybf)$. From Definition \ref{def:Ncal} this implies that  there exists some $\xhbf \in \Ncal(\Ccal(\Xcal,T),\ybf)$ such that $||\lambdaML \ybf - \xhbf|| < ||\lambdaML \ybf - \xML||$ however this would imply $L(\ybf;\xhbf,\hML) > L(\ybf;\xML,\hML)$ from (\ref{eq:LhoodEuc}) and hence we have a proof by contradiction.
\end{proof}

\section{PAM Detection For Real-Valued Fading Channels}\label{sec:PAMAlgorithm}

This section presents a low complexity algorithm for GLRT-optimal noncoherent PAM detection over real-valued channels. Practically, such channels arise in baseband transmission (eg.\ multi-level PCM), or in certain bandpass systems where phase and frequency are separately estimated by a phase-locked loop.

We first present a theorem that we will use to reduce the number of codewords that need to be examined, even beyond the limitations imposed by Property \ref{prop:xML}.  Note that in this real-valued channel case, the subspace of interest (defined by $\ybf$) actually reduces to a line, $\ybf \Rbb$.  The theorem implies that only a limited extent of the line needs to be searched; and that the extent depends on the largest value of $\ybf$.  We then propose a fast low-complexity iterative algorithm to perform the search. In the sequel, we will extend the algorithm to complex-valued channels. Later, we will directly incorporate the algorithm from this section into an extremely low complexity suboptimal algorithm for noncoherent detection over the more commonly encountered complex-valued channels.

\subsection{Limiting the Search Space} \label{sec:PAMLim}
\begin{theorem} \label{th:PAM}
For noncoherent detection of $M$-ary PAM codewords of length $T$ over a real-valued fading channel
\begin{equation*}
\abs{\lambdaML y_t} \leq M + T - 2
\end{equation*}
for all $t=1,2,\ldots,T$.
\end{theorem}
\begin{proof}
Define $n \defas \arg \max_t \cubr{\abs{ \lambdaML y_t}}$. Note that if $\abs{ \lambdaML y_n} \leq M$ then the theorem is satisfied.

Now, consider the alternative case when $\abs{\lambdaML y_n} > M$. Rearranging the GLRT-optimal channel estimate in (\ref{eq:hest}) gives $(\xML)'(\ybf-\hML\xML) = 0$ and hence
\begin{equation} \label{eq:dLdhML}
(\xML)'(\lambdaML \ybf-\xML) = 0.
\end{equation}
We will use this property to bound $\lambdaML$. Using (\ref{eq:LhoodEuc}) and the fact that $\Ccal(\Xcal,T)$ contains all possible sequences $\cubr{ \;\xbf\;|\;x_t \in \Xcal\;\forall t\;}$, the elements of $\xML$ can be determined on an element-wise basis as
\begin{equation} \label{eq:PAMxMLt}
\xho_t = \arg \min_{x \in \Xcal} \abs{ \lambdaML y_t-x }.
\end{equation}
For the case we are considering where $\abs{\lambdaML y_n} > M$, it follows that since the largest PAM constellation values are $\pm (M-1)$, that $\xho_n = \sgn \cubr{y_n}(M-1)$ where $\sgn$ is the signum function.

We now substitute (\ref{eq:PAMxMLt}) into (\ref{eq:dLdhML}) to bound $\lambdaML$, which gives
\begin{equation} \label{eq:PAMhsub}
\xho_n \br{ \lambdaML y_n-\xho_n}
= - \sum_{t\neq m} \xho_t \br{ \lambdaML y_t-\xho_t}.
\end{equation}
Now (\ref{eq:PAMxMLt}) and the symmetry of the PAM constellation implies that  $\sgn \cubr{\xho_t} = \sgn \cubr{ \lambdaML y_t}$. Moreover, since $\xho_t \in \cubr{\pm 1, \pm 3,\ldots,\pm (M-1) }$ it follows form the definition of $\lambdaML$ that $-1 < \lambdaML y_t - \xho_t < 1$ for all $\xho_t$ except $\pm(M-1)$. More generally, for all $\xho_t$, $ \sgn \cubr{\xho_t}(\lambdaML y_t-\xho_t) \geq -1$  and hence $\xho_t \br{ \lambdaML y_t-\xho_t} \geq -|\xho_t|$. Substituting this into (\ref{eq:PAMhsub}) gives
\begin{align*}
\xho_n \br{ \lambdaML y_n-\xho_n}
&\leq  \sum_{t\neq m} \abs{\xho_t}
\end{align*}
and hence
\begin{align*}
\abs{\lambdaML y_n}
&= \Bigg | \; \xho_n + \frac{1}{\xho_n} \sum_{t\neq n} \abs{\xho_t} \; \Bigg |\\
&\leq M-1 +\frac{(T-1)(M-1)}{M-1} = M + T - 2.
\end{align*}
Therefore, since $|\lambdaML y_t| \leq |\lambdaML y_n|$ for all $t$ the theorem is proved.
\end{proof}

\subsection{PAM Algorithm}\label{sec:PAMAlgorithmDetails}

In this Section we use Property \ref{prop:xML} and Theorem \ref{th:PAM} to develop a low-complexity algorithm for real-PAM detection. This algorithm reduces the number of codewords for which the decision metric is evaluated to order $MT$, which is much smaller than the set of all possible $M^T$ codewords that would be considered by an exhaustive search. Furthermore, we demonstrate how the algorithm can be implemented in an iterative manner so that the complexity is $O(T \log T)$.

\begin{figure}
\begin{center}
\includegraphics[width=0.5\linewidth]{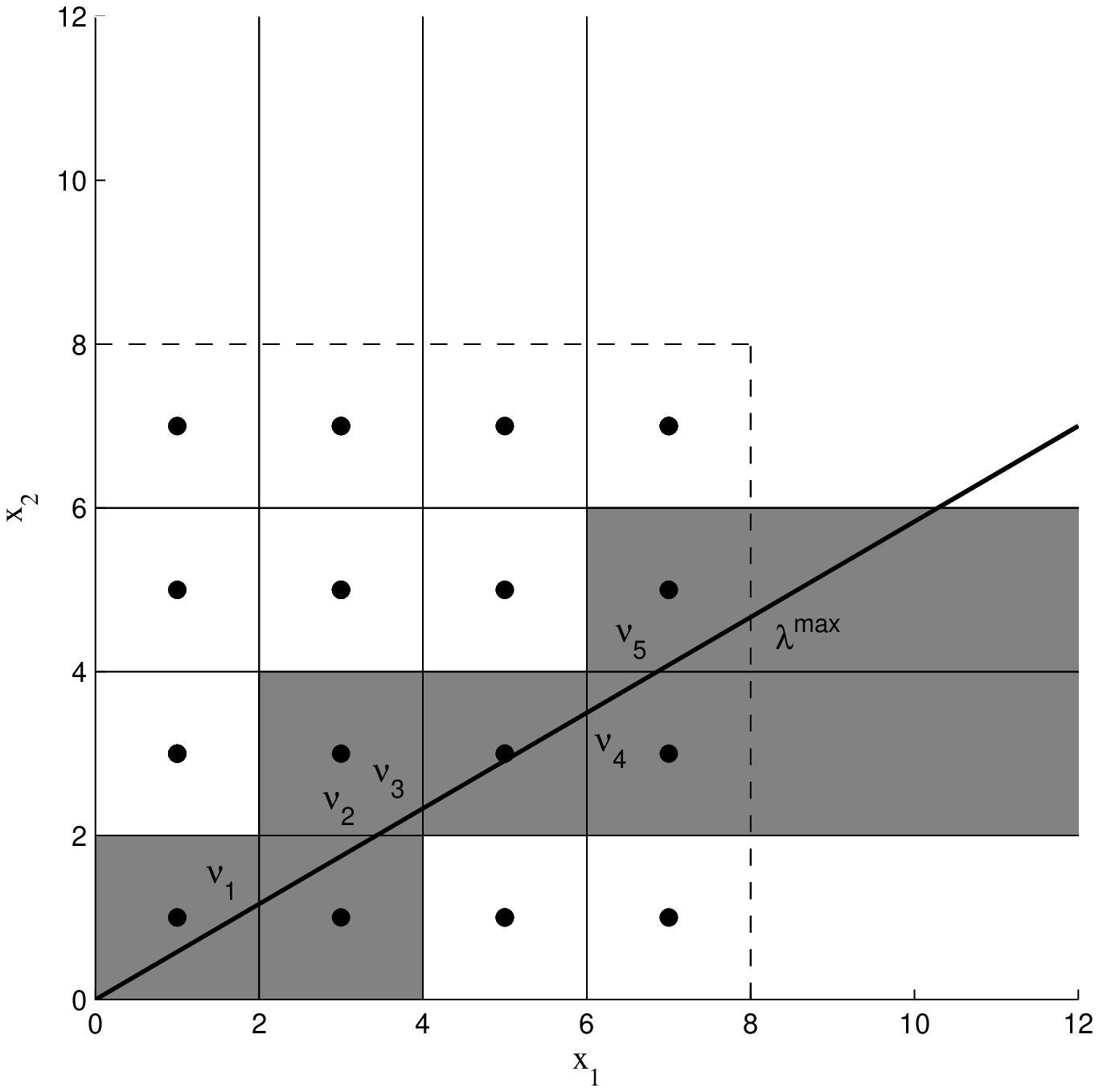}
\end{center}
\caption{Illustration of noncoherent detection of 8-ary PAM for $T=2$. The dots are all the (two dimensional) PAM codewords in the positive quarter-plane, and the angled line is $\ybf\Rbb$, for an example received codeword $\ybf$.  The shaded regions indicate the nearest neighbor regions of points which need to be searched.  That is, they are in $\Ncal(\Ccal(\Xcal,T),\ybf)$ (from Property \ref{prop:xML}), and they correspond to values of $\lambda$ less than $\lambdamax = (M+T-2)/\max_t \abs{y_t} = M/\max_t \abs{y_t}$ (from Theorem \ref{th:PAM}).}
\label{fig:pam_alg}
\end{figure}

Property \ref{prop:xML} implies that $\xML$ can be found by calculating the metric in (\ref{eq:x_ML}) for only those $\xhbf \in \Ncal(\Ccal(\Xcal,T),\ybf)$, \ie for only those $\xhbf \in \Ccal$  for which the line $\ybf\Rbb$ passes through its Euclidean nearest neighbor region. Furthermore, Theorem \ref{th:PAM} implies that only a finite segment of the line need be considered.  We have demonstrated such a search in Figure \ref{fig:pam_alg}, which shows the positive axes for 8-ary PAM with $T=2$, where the shaded regions indicate the nearest neighbor regions of the points which need to be searched. The specifics of the algorithm are as follows.

First, for ease of notation we modify the received codeword $\ybf$ by changing the signs of all negative elements in $\ybf$.  This will mean that the corresponding (modified) $\xML$ will now have all positive elements.  The true (original) GLRT estimate of $\xbf$ can be obtained by applying the reverse sign changes to $\xML$. Observe that we can do this without loss of generality since the PAM constellation is symmetric around zero.

\begin{defn} \label{def:Pxhat}
We define $P(\xhbf)$ to be the range of $\lambda$ such that $\xhbf$ is the nearest neighbor to $\lambda \ybf$, within the limits $0 < \lambda < \lambdamax \defas (M + T - 2)/\max_t \cubr{y_t}$ (where the limits are due to Theorem \ref{th:PAM} and the fact that all $x_t$ are greater than zero for the modified received codeword).  Formally,
\begin{align*}
P(\xhbf)
&\defas \cubr{ \lambda \; | \; \xhbf \in NN(\lambda \ybf),\;\lambda \in (0, \lambdamax) }.
\end{align*}
\end{defn}

Note that each non-empty $P(\xhbf)$ corresponds to a distinct interval of the line $\ybf\Rbb$.  The proposed algorithm proceeds by enumerating these non-empty $P(\xhbf)$'s, by first enumerating their boundary points along the line $\ybf\Rbb$.  We then sort the boundary points so that the decision metrics for the corresponding $\xhbf$ can be calculated in an iterative manner.

For real-PAM, the boundary values of $\lambda$ can be shown to be given by $\nu_{t,b} = \frac{b}{y_t}$ for all $t=1,\ldots,T$ and $b=2,4,\ldots,M-2$ such that $0 < \nu_{t,b} < \lambdamax$ (where the values of $b$ come from the regular boundaries in the positive half of the PAM constellation $\Xcal$).  We use $\Vcal_0$ to denote the set of all $(\nu_{t,b},t)$ pairs.

We then sort the elements of $\Vcal_0$ in ascending order of their $\nu_{t,b}$ value, and append the value  $(\lambdamax,0)$ to the end of the ordered set (since this is the outer boundary of the segment of the line $\ybf\Rbb$ which needs to be searched, according to Theorem \ref{th:PAM}; where the second element of the pair is arbitrarily set to 0 since it is not needed in the algorithm).  We denote the newly ordered set by $\Vcal$, and index it by $k$, (\ie we denote its $k$th element by $(\nu_k\,,\,t_k)$).  These ordered values are shown on the example case of Figure \ref{fig:pam_alg}, where the values of $\nu_k$ denote the distance along the line $\ybf\Rbb$ where the line crosses from one nearest neighbor region into the next.

We now show how $\Vcal$ can be used to enumerate the codewords which need to be searched, and show how to calculate the corresponding decision metrics in an iterative manner.  The algorithm can be visualized geometrically as searching along a segment of the line $\ybf\Rbb$, by iterating in $k$. Whenever the line crosses from the nearest neighbor region of one lattice point (codeword) to the nearest neighbor region of another, we calculate the metric for the new lattice point.  Note that in this context, the value of $t_k$ indicates the dimension of the boundary that is going to be crossed (in the $T$-dimensional space) when {\em leaving} the $k$th segment of the line.

The iterative search starts with the codeword $\xhbf^{(1)} = \onebf \defas [\;1,1,\ldots,1]'$; which has a corresponding decision metric $L(\onebf) = (\onebf'\ybf)^2/\magn{\onebf}^2$, where $L(\xhbf) \defas (\xhbf'\ybf)^2/\magn{\xhbf}^2$ is the likelihood function in (\ref{eq:x_ML}).  We will use the symbol $\hat{\lambda}$ as a marker for the most likely codeword, and we initially set it to $\hat{\lambda} = \nu_1/2$ (\ie during the iteration process, $\hat{\lambda}$ will be updated whenever a codeword is found to have a higher likelihood than any previously searched codeword, and the value of $\hat{\lambda}$ will be chosen such that $NN(\hat{\lambda})$ gives the new codeword).

The iteration proceeds by noting that each time a nearest neighbor boundary is crossed, only one element of the $T$-dimensional nearest neighbor codeword vector changes (since for real-PAM, the boundaries are straight lines, orthogonal to one of the dimensions, and parallel to all the others).  Therefore the $k$th codeword which needs to be considered, is calculated from the $(k-1)$th codeword, on an element-wise basis as follows:
\begin{equation}
\xh^{(k)}_{p} = \left\{\begin{array}{ll}\xh^{(k-1)}_{p}, & \text{for}\;p\neq t_{k-1}\\
 \xh^{(k)}_{p} + 1, & \text{for}\;p = t_{k-1}. \end{array}\right.
\label{eq:pamxupdate}
\end{equation}
We define $\alpha_k \defas (\xhbf^{(k)})'\ybf$ and $\beta_k \defas \magn{\xhbf^{(k)}}^2$, and hence $L(\xhbf^{(k)}) = \alpha_k^2/\beta_k$ is the decision metric for the $k$th codeword considered. The values $\alpha_k$ and $\beta_k$ are calculated iteratively as follows,
\begin{align}
\alpha_{k} &= \alpha_{k-1} + 2y_{t_{k-1}}\label{eq:pamaupdate}\\
\beta_{k} &= \beta_{k-1}  + 4\xh_{t_{k-1}} + 4.
\label{eq:pambupdate}
\end{align}

If $L(\xhbf^{(k)})$ improves on the previous best codeword estimate then we update $\hat{\lambda}$ in the interior of $P(\xhbf^{(k)})$, by setting $\hat{\lambda} = (\nu_k + \nu_{k-1})/2$.

Once all segments of the line have been searched, we have $\xML = NN(\hat{\lambda} \ybf)$.  Pseudo-code for the algorithm is given in Table \ref{tab:algorithm}.

The complexity of the algorithm is a function of the number of intersection points $\nu_{t,b}$, \emph{i.e.}, $N_I \defas \abs{\Vcal_0}$, where $\abs{\cdot}$ denotes set cardinality.  $N_I$ is upper bounded by $(M/2-1)T$, however in general it will be much less than this due to the restricted line search implied by Theorem \ref{th:PAM}, as shown by simulation in Section \ref{sec:Sims}.  The sorting of $\Vcal_0$ can be performed using standard sorting techniques in $O(N_I \log N_I)$ \cite{Cormen01}. The updates (\ref{eq:pamxupdate}), (\ref{eq:pamaupdate}) and (\ref{eq:pambupdate}) have complexity $O(1)$, and the final calculation of $\xML$ is of order $T$. Thus the overall complexity is dominated by the sorting operation, and hence the complexity of the algorithm is of order $O(T \log T)$; a significant improvement compared with an exhaustive search over all $M^T$ possible codewords in the codebook $\Ccal(\Xcal,T)$.

\section{GLRT-Optimal QAM Detection For Complex-Valued Fading Channels}

This section presents a low complexity algorithm for GLRT-optimal noncoherent QAM detection over complex-valued  fading channels. Similarly to the the real-PAM case, we first present a theorem that we will use to reduce the number of codewords that need to be examined, beyond the limitations imposed by Property \ref{prop:xML}. In the complex-valued channel case, the subspace of interest is the plane $\Yubbf \Rbb^2$, where $\Yubbf$ was defined in (\ref{eq:Ydef}).  The theorem implies that only a limited extent of the plane needs to be searched; and that the extent depends on the largest element in $\ybf$.  We then propose a fast low-complexity algorithm to perform the search for QAM. We also show how PAM detection over complex-valued channels can be viewed as a special case of the QAM algorithm.

\subsection{Limiting the Search Space}
\begin{theorem} \label{th:QAM}
For noncoherent detection of $M^2$-ary QAM codewords of length $T$ over a complex-valued fading channel
\begin{align*}
\abs{\Real{\lambdaML y_t}} &\leq  M + 2T - 2, \;\;\;\text{and} \\
\abs{\Imag{\lambdaML y_t}} &\leq  M + 2T - 2,
\end{align*}
for all $t = 1,2,\ldots,T$.
\end{theorem}
\begin{proof}
Define the point $\vbf \defas \lambdaML \ybf$, along with its corresponding real-valued representation $\vubbf$, as in (\ref{eq:c2r}).  Also define $n \defas \arg \max_t \cubr{\abs{ v_t}}$. Note that if $\abs{ v_n } \leq M$ then $\abs{\Real{\lambdaML y_t}} \leq M$ and $\abs{\Imag{\lambdaML y_t}} \leq M$ for all $t$ and the theorem is satisfied.

Now, consider the alternative case when $\abs{v_n} > M$. Similarly to the real-PAM case, rearranging the GLRT-optimal channel estimate in (\ref{eq:hest}) gives $(\xML)^\dag(\ybf-\hML\xML) = 0$ and hence
\begin{equation} \label{eq:dLdhMLcomplex}
(\xML)^\dag(\lambdaML \ybf-\xML) = 0.
\end{equation}
It follows that
\begin{equation*}
\lambdaML = \frac{\magn{\xML}^2}{\abs{\ybf^\dag \xML }^2} \ybf^\dag \xML.
\end{equation*}
Combining this with the the fact that for any vector $\ubf \in \Cbb^T$, the real-valued representation of the complex scalar $\ybf^{\dag} \ubf$ is $\Yubbf' \uubbf$, we obtain the real-valued representation of $\lambdaML$ as
\begin{equation*}
\lambdabfML =  \frac{\magn{\xubbf}^2}{\magn{\Yubbf' \xubbf}^2} \Yubbf' \xubbf
\end{equation*}
and therefore
\begin{equation*}
\vubbf = \Yubbf \lambdabfML =  \frac{\magn{\xubbf}^2}{\magn{\Yubbf' \xubbf}^2} \Yubbf \Yubbf' \xubbf.
\end{equation*}
It follows that $\vubbf' \xubbf = \magn{\xubbf}^2$, \emph{i.e.}
\begin{equation} \label{eq:dLdhMLC}
(\xubML)'(\vubbf - \xubML) = 0.
\end{equation}
Using (\ref{eq:LhoodEuc}) and the fact that $\Ccal(\Xcal,T)$ contains all possible sequences $\cubr{ \;\xbf\;|\;x_t \in \Xcal\;\forall t\;}$, the elements of $\xML$ can be determined on an element-wise basis as
\begin{align}\label{eq:xoptQAM}
\xhub^{\text{opt}}_t &= \arg\min_{\xub \in \Xcal'} \abs{ \vub_t-\xub}
\end{align}
for all $t=1,\ldots,2T$ where we recall from Section \ref{sec:SigModel} that $\Xcal' = \cubr{\pm 1,\pm 3,\ldots,\pm(M-1)}$.
We now substitute (\ref{eq:xoptQAM}) into (\ref{eq:dLdhMLC}) which gives
\begin{equation}
\xhubo_n \br{ \vub_n-\xhubo_n}
= - \sum_{t\neq n} \xhubo_t \br{ \vub_t-\xhubo_t}.
\end{equation}
This is similar to (\ref{eq:PAMhsub}) in the proof of Theorem 1. By following through the subsequent steps in the proof of Theorem 1, and keeping in mind that the dimensions of the vectors are now of dimension $2T$, we obtain $\abs{\vub_n} \leq  M + 2T - 2$ which implies that $\abs{\Real{\lambdaML y_t}} < M + 2T - 2$ and $\abs{\Imag{\lambdaML y_t}} < M + 2T - 2$ for all $t=1,\ldots,T$.
\end{proof}

\subsection{QAM Algorithm}\label{sec:QAM_algorithm}

\begin{figure}
\begin{center}
\includegraphics[width=0.7\linewidth]{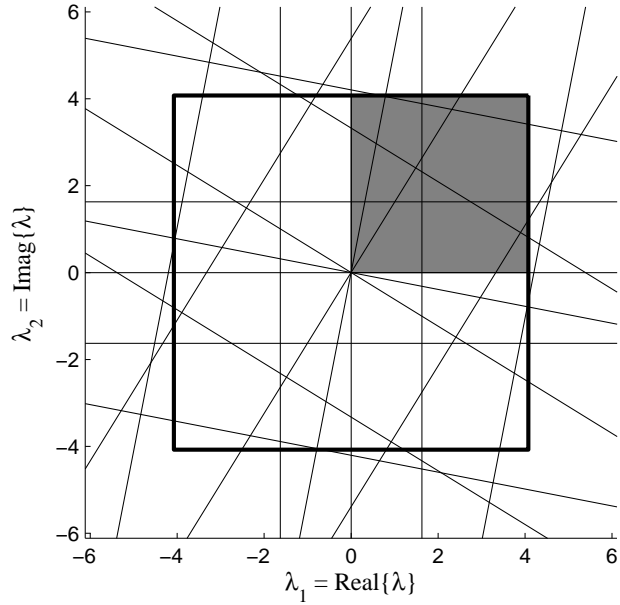}
\end{center}
\caption{Plot of partitions $P(\xhbf)$ on the $\Rbb^2$ plane for $16$-ary QAM detection of a sequence of length $T=3$ for the received vector $\ybf = [\;-0.1076 - 0.4728i,\;  -0.7002 - 0.0968i,\;  -1.1228 + 0.4955i\;]$. The bold square corresponds to the search boundary $\Scal$. }
\label{fig:qam_lines}
\end{figure}

In this Section we use Property \ref{prop:xML} and Theorem \ref{th:QAM} to develop a low-complexity algorithm for QAM detection. Property~\ref{prop:xML} implies that $\xML$ can be found by calculating the metric in (\ref{eq:x_ML}) for only those $\hat{\xubbf} \in \Ncal(\Ccal(\Xcal,T),\Yubbf)$, \ie for only those $\hat{\xubbf} \in \Ccal$  for which the plane $\Yubbf\Rbb^2$ passes through its Euclidean nearest neighbor region. Furthermore, Theorem \ref{th:QAM} implies that only a finite region of the plane need be considered.  Conceptually, this is a direct extension of the real-PAM case shown in Figure \ref{fig:pam_alg} (considered previously).  The difference being that Figure \ref{fig:pam_alg} shows the line $\ybf\Rbb$, but we now have a plane $\Yubbf\Rbb^2$.  Also the number of orthogonal dimensions doubles when considering complex-valued channels.  We demonstrated this complex-valued channel QAM case in Figure \ref{fig:qam_lines} which is a two dimensional plot in the plane $\Yubbf\Rbb^2$.  The parallel lines (at various angles) are the boundaries arising from the QAM constellation, and the shaded region indicates the nearest neighbor regions of codewords which need to be searched. The QAM search algorithm we present here, follows the same principles as the real-PAM algorithm of Section \ref{sec:PAMAlgorithmDetails}, where instead of working with boundary points of line segments, we need to work with boundary edges of planar regions. The specifics of the algorithm are as follows.

First, for ease of notation we modify the received codeword $\ybf$ by multiplying it by the complex scalar $y_m^*/\abs{y_m}$, where $m = \arg \max_t \abs{y_t}$. This will mean that the $m$th element of $\ybf$ will be real-valued and positive.  The true (original) GLRT-optimal estimate of the channel can be be obtained by applying the reverse phase rotation to $\hML$, while the optimality of the new GLRT-optimal codeword estimate is unaffected.

Hence Theorem  \ref{th:QAM} implies that the search over the plane $\Yubbf \lambdabf$ is reduced to the segment of the plane for which $\abs{\lambda_1},\abs{\lambda_2} < \lambdamax$ where $\lambdamax \defas (M+2T-2)/\abs{y_m}$. Furthermore, as discussed in Section \ref{sec:Detection} because of the $\pi/2$ phase ambiguity in square QAM constellations, there are four GLRT-optimal inverse channel estimates $\pm \lambdaML, \pm i \lambdaML$ (with corresponding phase ambiguous GLRT-optimal codeword estimates). Hence, we only need to consider the square region of the plane
\begin{equation}\label{def:Scal}
\Scal = \cubr{\; \lambdabf\;|\;\lambda_1 \in (0,\lambdamax),\; \lambda_2 \in [0,\lambdamax)\;}
\end{equation}
since exactly one of $\pm \lambdaML, \pm i \lambdaML$ will exist in this region of the plane.  Note that $\Scal$ is the shaded region in Figure \ref{fig:qam_lines} (mentioned previously).

Similarly to the real-PAM case we make the following definition.
\begin{defn} \label{def:Qxhat}
We define $P(\xhubf)$ to be the range of $\lambdabf\in \Scal$ such that $\xhubf$ is the nearest neighbor to $\Yubbf \lambdabf$.  Formally,
\begin{align*}
P(\xhubf)
\defas \cubr{ \lambdabf \; | \;\xhubf \in NN(\Yubbf \lambdabf ),\;\lambdabf \in \Scal\;}.
\end{align*}
\end{defn}

Note that each non-empty $P(\xhubf)$ corresponds to a distinct region of the plane $\Yubbf\Rbb^2$.  The proposed algorithm proceeds by enumerating these non-empty $P(\xhubf)$'s, by first enumerating their boundary vertices in the plane.  These vertices are found by calculating the intersection of all the constellation-point boundary lines in the plane (e.g.\ as shown in Figure \ref{fig:qam_lines}).  The vertices are then used to calculate an interior-point inside each of the nearest neighbor regions in the shaded square $\Scal$.  The respective nearest neighbor codeword is calculated for each interior-point, and then it is only these points for which the likelihood metrics are calculated. Clearly, this is a significantly reduced search space compared with the space of all possible codewords.

For QAM the vertices of the nearest-neighbor regions in the plane $\Yubbf\Rbb^2$ can be found by first noting that, since $\xhubo_t$ can be given in on an element-wise basis as in (\ref{eq:xoptQAM}), $P(\xhubf)$ can be written as
\begin{align*}
P(\xhubf)
&\defas \bigcap_{t=1}^{2T} \{ \; \lambdabf \; | \; \xub_t = \arg\min_{\xub \in {\Xcal'} } \abs{(\Yubbf \lambdabf)_{t}-\xub }, \; \lambdabf \in \Scal\}
\end{align*}
where $(\Yubbf \lambdabf)_{t}$ is the $t$th element of $\Yubbf \lambdabf$ and we recall that $\Xcal' = \cubr{\pm1,\pm3,\ldots,\pm (M-1)}$. This can be written as the feasible region for the set of linear inequalities corresponding to the nearest neighbor region boundaries in $\Xcal'$ for each element of $\xhub_t$, as
\begin{align*}
P(\xhubf)
= \bigcap_{t=1}^{2T} \cubr{ \; \lambdabf \; | \; l(\xhub_t) \leq (\Yubbf \lambdabf)_t \leq u(\xhub_t), \; \lambdabf \in \Scal }
\end{align*}
where $l(\xhub_t)$ and $u(\xhub_t)$ are the upper and lower nearest neighbor boundaries in the constellation $\Xcal'$. For $t \notin \cubr{2m,2m-1}$ they take on values in the set $\cubr{0, \pm 2,\ldots, \pm (M-2), \pm \infty}$. For $t \in \cubr{2m,2m-1}$ we must consider intersections with the boundary of $\Scal$, and therefore in this case $l(\xhub_t)$ and $u(\xhub_t)$ take on values in the set $\cubr{0, \pm 2,\ldots, \pm (M-2), \pm (M+2T-2)}$.

By including the square boundary of the region $\Scal$, all non-empty $P(\xhubf)$ are closed simply connected sets on the plane $\Rbb^2$. Therefore, since $P(\xhubf)$ is formed from linear inequalities it is a convex polygon in $\Rbb^2$. For each $P(\xhubf)$, denote $B(\xhubf)$ as its polygonal boundary and $V(\xhubf)$ as the vertices of the polygon.

We now propose a method that enumerates all the vertices $V(\xhubf)$ for all non-empty $P(\xhubf)$, and then uses these vertices to generate a point in the interior of all $P(\xhubf)$, which is then used to obtain a unique codeword via finding the nearest neighbor codeword to that point. Consider the set of points $\cubr{\; \nubf \pm \epsilon \mubf \; | \nubf \in V(\xhubf) }$. If $\mubf$ is some vector that is not parallel to any side of the polygon $P(\xhubf)$, and if $\epsilon$ is chosen sufficiently small, then at least one point in this set will be in the interior of $P(\xubbf)$. Since the received symbol is subject to AWGN, and is therefore irrational with probability one, it follows that the arbitrary choice of $\mubf \defas [\; 1 \; 1 \;]'$ will almost surely guarantee this, given that $\epsilon > 0$ is chosen sufficiently small. In practice, simply setting $\epsilon$ to some small positive constant will be sufficient to ensure that a point in the interior of $P(\xhubf)$ is enumerated. However, in Appendix \ref{app:exactoptimality} we present a technique to perform this in a strictly optimal fashion with complexity per vertex of $O(T)$.

Since the vertices are shared by adjacent $P(\xhubf)$, each vertex is only required to be enumerated once. We define the set of all vertices within or on the boundary of $\Scal$ as $\Vcal = \cubr{\;\nubf\;| \;\nubf \in V(\xhubf),\;P(\xhubf) \neq \emptyset \; } $. The set $\Vcal$ can be enumerated as the the intersections of the lines  $\Yub_{t,1}\nu_1 + \Yub_{t,2} \nu_2 = b$ and $\Yub_{t',1}\nu_1 + \Yub_{t',2} \nu_2 = b'$, for all pairs of  $t,t'$
and for all nearest neighbor boundaries $b,b'$ in $\Xcal'$. That is
\begin{equation} \label{eq:Vcal_solve}
  \sqbr{
  \begin{matrix} \nu_1 \\ \nu_2 \end{matrix}
  }
  =
  \sqbr{
  \begin{matrix} \Yub_{t,1} &\Yub_{t,1} \\ \Yub_{t,2} &\Yub_{t,2}\end{matrix}
  }^{-1}
  \sqbr{
  \begin{matrix} b \\ b' \end{matrix}
  }
\end{equation}
for all $t=1,2,\ldots,2T-1$, $t' = t+1,t+2,\ldots,2T$, and $b,b' \in \Bcal(t)$, where $\Bcal(t) \defas \cubr{0, \pm 2,\ldots, \pm (M-2)}$ if $t \notin \cubr{2m,2m-1}$ and for symbol indices $t \in \cubr{2m,2m-1}$ where we consider the square boundary $\Bcal(t) \defas \cubr{0, \pm 2,\ldots, \pm (M-2), \pm (M+2T-2)}$.

To enumerate a point in each $P(\xhubf)$, for each vertex $\nubf$ enumerated we calculate the points on the plane $\lambdabf^+ \defas \nubf + \epsilon \mubf$, and $\lambdabf^- \defas  \nubf + \epsilon \mubf$. Then for each of these two points, if it is in the square $\Scal$, we calculate the corresponding codewords $NN(\lambdabf^+ \Yubbf)$ and/or $NN(\lambdabf^- \Yubbf)$ and the decision metrics in (\ref{eq:x_ML}).

Pseudo-code is provided in Table \ref{tab:qam_algorithm}.

The complexity of the algorithm is a function of the number of codewords examined, $N_C$, which is in turn a function of the number of vertices calculated. The number of vertices calculated in (\ref{eq:Vcal_solve}) corresponding to the intersections between lines in  where $b,b'$ is a boundary of $\Xcal'$ and both $b$ and $b'$ are non-zero is $T(2T-1)[ (M-1)^2-1 ]$; for which at most two codewords are generated for a quarter of these intersections. For the intersections of the boundaries of $\Xcal'$ and the square $\Scal$ there are $2(2T-2)(M-1)^2$ intersections, which for one quarter of these intersections one codeword is generated. For the vertices at $(0,0)$ and $(\lambdamax,\lambdamax)$ one codeword is generated. Hence the total number of codewords examined is at most
\begin{equation} \label{eq:NCQAM}
N_C \leq \frac{T(2T-1)}{2} [ (M-1)^2-1 ] + (2T-2)(M-1)^2 + 2.
\end{equation}
Since the complexity of each codeword and decision metric calculation is of order $T$ then the overall complexity is of order $M^2 T^3$ (which is linear in the constellation size $M^2$) a significant improvement over an exhaustive search over all $M^{2T}$ possible codewords in the codebook $\Ccal(\Xcal,T)$.

A further reduction in computational expense, without any loss in optimality, can be achieved by enumerating only one out of each set of four phase ambiguous vertices. The technique is not presented here due to space constraints, however the number of non-zero vertices examined is reduced by a factor of 4 and $1/3$ of the matrix inverse calculations in (\ref{eq:Vcal_solve}) are avoided.

\subsection{PAM Over Complex Channels}\label{sec:PAMCalgorithm}

PAM detection over complex fading channels can be viewed as a special case of complex-channel QAM, where there is zero imaginary component in the constellation. In this case, the search over the plane $\Yubbf \Rbb^2$ can be restricted by extending the proof of Theorem \ref{th:QAM}. To do this, we note that the condition in (\ref{eq:dLdhMLcomplex}) holds, which implies that $(\xML)'( \Real{\lambdaML \ybf}-\xML) = 0$ since $\xML$ is always real-valued. The rest of the proof follows to give the result that $\abs{ \Real{\lambdaML y_t}} \leq M + T - 2$. This fact combined with Property \ref{prop:xML} and the $\pi$ phase ambiguity of PAM constellations, implies that we only consider codewords $\xhubf = NN(\Yubbf \lambdabf)$ for $\lambdabf$ in the region $\Scal = \cubr{ \;\lambdabf\;|\;0 < \lambda_1 < \lambdamax = (M + T - 2)/\abs{y_m}\;}$.

The specifics of the $M$-ary PAM algorithm are the same as for the $M^2$-ary QAM case, with the exception that the calculation of (\ref{eq:Vcal_solve}) to obtain the vertices in the interior of the (\ref{eq:Vcal_solve}) is only performed for all $t=1,3,\ldots,2T-1$, $t' = t+2,t+4,\ldots,2T$, and $b,b' \in \Bcal(t)$, where $\Bcal(t) \defas \cubr{0, \pm 2,\ldots, \pm (M-2)}$ if $t \neq 2m-1$ and for $\Bcal(t) \defas \cubr{0, \pm 2,\ldots, \pm (M-2), M+T-2}$ if $t = 2m-1$.

The total number of codewords searched can be shown to be at most
\begin{equation} \label{eq:NCPAM}
N_C \leq \frac{T(T-1)}{2} [ (M-1)^2-1 ] + (T-1)(M-1) + 1.
\end{equation}
Since the complexity of each codeword and decision metric calculation is of order $T$ then the overall complexity is $O(T^3)$.  In the following section, we will see that a simple suboptimal approach can achieve even lower complexity with near-optimal performance.

\section{Suboptimal Algorithms for Complex-Valued Fading Channels} \label{sec:Subopt}
In this section, we propose even lower complexity suboptimal algorithms for detection of QAM and PAM over complex-valued fading channels. We directly use the GLRT-optimal algorithm for real-PAM from Section \ref{sec:PAMAlgorithm} as the basis for the algorithms.

\subsection{Suboptimal PAM algorithm} \label{sec:SubPAMCalgorithm}
Since for PAM constellations, all constellation points lie along the real line in the complex plane, a suboptimal phase estimation technique combined with our GLRT-optimal algorithm for real-valued fading channels should be sufficient to provide near-optimal performance. This effectively reduces the search over the whole plane $\Yubbf \Rbb^2$ for the GLRT-optimal case, to a search over a single line at the given estimated phase angle.

We use the power-law estimator \cite{Moen94} which, for constellations exhibiting a rotational symmetry of $\pi$ radians, is simply
\begin{equation}
\phih_{\text{PL}} \defas \frac{1}{2}  \angle {\sum_{t=1}^T y_t^2}
\end{equation}
where $\angle$ refers to the complex argument. Detection is performed by first rotating the received codeword $\ybf$ according to this estimate, and then detecting $\operatorname{Re} \{ e^{-j\phih_{\text{PL}} }\ybf \}$ using the GLRT-optimal algorithm of PAM over a real-valued fading channel.

\subsection{Suboptimal QAM algorithm} \label{sec:SubQAMalgorithm}

Here we propose a suboptimal algorithm, which reduces the overall algorithmic complexity to $O(T^2 \log T)$ by using $O(T)$ instances of the PAM detection algorithm presented in Section \ref{sec:PAMAlgorithm}. Instead of enumerating the intersections of lines on the $(\lambda_1,\lambda_2)$-plane, as we did in Section \ref{sec:QAM_algorithm}, here we propose to use a modified version of the nearest-neighbor real-PAM line-search algorithm for $L$ lines of the type presented in Section \ref{sec:PAMAlgorithm}. We generate these lines emanating from the origin into $\Scal$ (the shaded region in Figure \ref{fig:qam_lines}), evenly spaced in angle.  Of course, this does not guarantee that we fully enumerate $\Ncal(\Ccal(\Xcal),T)$ since a finite number of radiating lines can not completely cover a plane, however, we will see by simulation in Section \ref{sec:QAM_algorithm} that the performance is close to the optimal.

As in the optimal case, we multiply $\ybf$ by $y_m^*/\abs{y_m}$ so that $y_m$ will be real-valued and positive.  In this suboptimal QAM case, this implies that we only examine points on the plane $\Yubbf \lambdabf$ for $\lambdabf = [\;\lambda_1\;\lambda_2\;]'$ satisfying $0 < \abs{\lambda_1},\abs{\lambda_2} < \lambdamax_0$, where $\lambdamax_0 \defas (M+2T-2)/\abs{y_m}$.

The $L$ directions of the lines with respect to the direction of positive $\lambda_1$ have angles $\Phi \defas \cubr{ \phi_1,\ldots,\phi_L}$ where $\phi_{\ell} = ({\ell}-1)\pi/(2L)$. For each angle $\phi_{\ell}$, we perform a nearest neighbor line search for the line with basis vector $\yubbf_{\ell} [\;\cos \phi_{\ell} \; \sin \phi_{\ell} \; ]$, as proposed in the suboptimal PAM algorithm in Section \ref{sec:SubPAMCalgorithm}. The search is performed for the segment of the line $\lambda \yubbf_{\ell}$ where $\lambda \in \Rbb$ and  $0 < \lambda < \lambdamax_{\ell}$ where $\lambdamax_l \defas \lambdamax_0 / \max \cubr{ \cos \phi_{\ell}, \sin \phi_{\ell} }$.  In this case the lines searches are performed for blocks of length $2T$.

There is of course a modification required to update the codeword metrics in terms of complex numbers.  The first line search performed is for $\phi_1 = 0$, and hence the line search is over $\lambda \yubbf_1 = \lambda \yubbf$. In this case the intervals of the line $P(\xhubf)$ are defined as,
\begin{align*}
P(\xhubf)
&\defas \cubr{ \lambda \; | \; \xhubf \in NN(\lambda \yubbf_l),\; \lambda \in (0,\lambdamax_l)}.
\end{align*}
Hence the algorithm works by enumerating and calculating the metric for all $\xhubf \in \Ccal(\Xcal',2T)$ for which $P(\xhubf)$ is non-empty.

In this case the set $\Vcal_0$ of boundary points of the regions $P(\xhubf)$ is enumerated by calculating $\nu_{t,b} = b /|\yub_t|$ for all $t=1,\ldots,2T$ and $b=2,4,\ldots,M-2$, (which are the nearest neighbor boundaries in the positive half of the constellation $\Xcal'$), and storing only those values of $(\nu_{t,b},t)$ such that $\nu_{t,b} < \lambdamax_{\ell}$. The set of ordered boundary points $\Vcal$ is again obtained by sorting, and $(\lambdamax_{\ell},0)$ is appended to $\Vcal$ as the extent of the search. Recall that $(\nu_k,t_k)$ are the $k$th elements of $\Vcal$.

The search through the codewords is initialized to the first codeword for the which the line segment passes through, which is given by $\xhubf^{(1)} = \subbf$ where $\subbf \defas \sgn \cubr{ \yubbf }$. The likelihood update variables are initialized to $\alpha = (\xhbf^{(1)})^{\dag}\ybf$ and $\beta = \| \xhubf^{(1)} \|^2$. To regenerate the optimal codeword, the values of $\lambda$ and $\phi$ are initialized to $\lambda = \nu_1/2$ and $\phi = \phi_{1}=0$.

The $(k+1)$th codeword considered, $\xhbf^{(k+1)}$, is calculated from the $k$th codeword as
\begin{equation}
\xhub^{(k+1)}_{t_k} = \xhub^{(k)}_{t_k} + 2\sub_{t_k}.
\end{equation}
To update the decision metric we define $\alpha_k \defas (\xhbf^{(k)})^\dag\ybf$ and $\beta_k \defas \| \xhubf^{(k)} \|^2$, and hence $L(\xhbf^{(k)}) = |\alpha_k|^2/\beta_k$ is the decision metric for the $k$th codeword considered. The values $\alpha_k$ are updated as follows,
If $t_k$ is odd, then $\alpha_k$ is updated as
\begin{equation} 
\alpha_{k} =
\begin{cases}
 \alpha_{k-1} + 2 \sub_{t_{k-1}} y_{(t_{k-1}+1)/2}, & t_{k-1} \text{ odd}\\
 \alpha_{k-1} - 2i\sub_{t_{k-1}} y_{t_{k-1}/2}, & t_{k-1} \text{even}.
\end{cases}
\end{equation}
The values of $\beta_k$ are updated according to
\begin{equation} 
\beta_{k} = \beta_{k-1}  + 4 \sub_{t_{k-1}} \xhub_{t_{k-1}} + 4.
\end{equation}
If $L(\xhbf^{(k)}) = |\alpha_k|^2/\beta_k$ improves on the best codeword estimate then we store $\lambda = (\nu_k + \nu_{k-1})/2$ and $\phi = \phi_{\ell}$.

To start the next line search, $\ybf$ is multiplied by $e^{\frac{j \pi}{2L}}$ and the line search is then performed again for the new value of $\yubbf$. When all line searches have been performed, we calculate $\xML = NN(\lambda e^{j \phi} \ybf )$ for the original $\ybf$.

Pseudo-code is provided in Table \ref{tab:QAMSub}.

The significantly reduced algorithmic complexity compared to the GLRT-optimal algorithm is governed by the number of line searches and the complexity of each line search. Since there are $L$ phases, each performing a version of the real-PAM line-search algorithm of Section \ref{sec:PAMAlgorithm} for the case $M$-ary PAM detection of $2T$ symbols. Thus $N_C \leq L(2T(M/2-1)+1)$. From Section \ref{sec:QAM_algorithm} we have noted that the number of codewords in $\Ncal(\Ccal(\Xcal,T),\ybf)$ is of order $M^2 T^2$ and thus $L$ must be $O(T)$ for it to be possible that the majority of $\Ncal(\Ccal(\Xcal,T),\ybf)$ is enumerated. Hence, if $L$ is increased proportionally to $T$, the overall complexity of the algorithm is $O(T^2 \log T)$.   Note that however, the improved computational performance of the algorithm is largely due to being able to choose $L$ small, which corresponds to avoiding examining a significant number of the $\xhubf$ with associated $P(\xhubf)$ being so small as to imply that $\xhubf$ is not relatively close in angle to the plane $\Yubbf \Rbb^2$. We will see via simulation in Section \ref{sec:Sims} that small $L$ (\eg $L=4$ for $T=7$ 16-QAM detection) can achieve near-optimal performance.

\section{Simulation Results} \label{sec:Sims}

We now present simulation results to demonstrate the performance of the new PAM and QAM noncoherent reduced search lattice-decoding algorithms. Simulations are performed to obtain the codeword error rate (CER) as a function of SNR for noncoherent detection of 8-ary PAM and 16-ary square QAM. For both case, the simulations are performed for codeword lengths of $T=3$ and $7$ over a block Rayleigh fading channel where $h$ is i.i.d. circularly symmetric complex Gaussian with unit variance. We have assumed that the phase ambiguities have been removed within each codeword, (for example, by the use of differential encoding \cite{Weber78}).

\begin{figure}
\begin{center}
\includegraphics[width=0.7\linewidth]{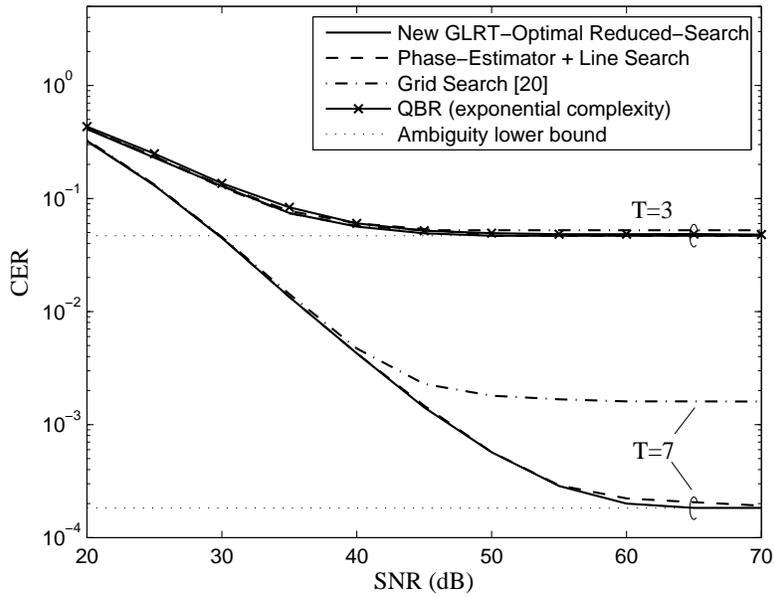}
\end{center}
\caption{\small Plot of Codeword Error Rate (CER) as a function of SNR for an 8-ary PAM system.}
\label{fig:pam_bler}
\end{figure}

Figure \ref{fig:pam_bler} presents results for 8-ary PAM for the GLRT-optimal plane search algorithm from Section \ref{sec:PAMCalgorithm} and the suboptimal phase-estimator plus line-search algorithm from Section \ref{sec:SubPAMCalgorithm}. We also compare with the suboptimal grid-search algorithm proposed in \cite{Motedayen03} and the quantization based receiver proposed in \cite{Warrier02}. For the grid-search algorithm we use uniformly spaced channel phase estimates and the channel attenuation estimates are chosen uniformly from the CDF of the Rayleigh fading channel distribution. For fairness the number of channel attenuation estimates is adjusted so that the total number of channel estimates was kept equal to the maximum number of codeword estimates that potentially could be produced by our GLRT-optimal algorithm. Best results are obtained for choosing the channel phase estimates as $0$ and $\pi/2$, and hence the $k$th channel amplitude estimate is given by $|\hh^{(k)}|^2 = -  \log ( 1- k/(1+ \lceil N_C/L \rceil ))$. For the quantization-based receiver (QBR) considered in \cite{Warrier02}, all possible sequences of (positive) amplitude levels are produced, and the sign of each symbol is then determined by symbol-wise coherent detection using uniformly spaced channel phase estimates (a channel amplitude estimate is not required since the signal amplitude is assumed known). For QBR, we again use the channel phase estimates $0$ and $\pi/2$.

\begin{figure}
\begin{center}
\includegraphics[width=0.75\linewidth]{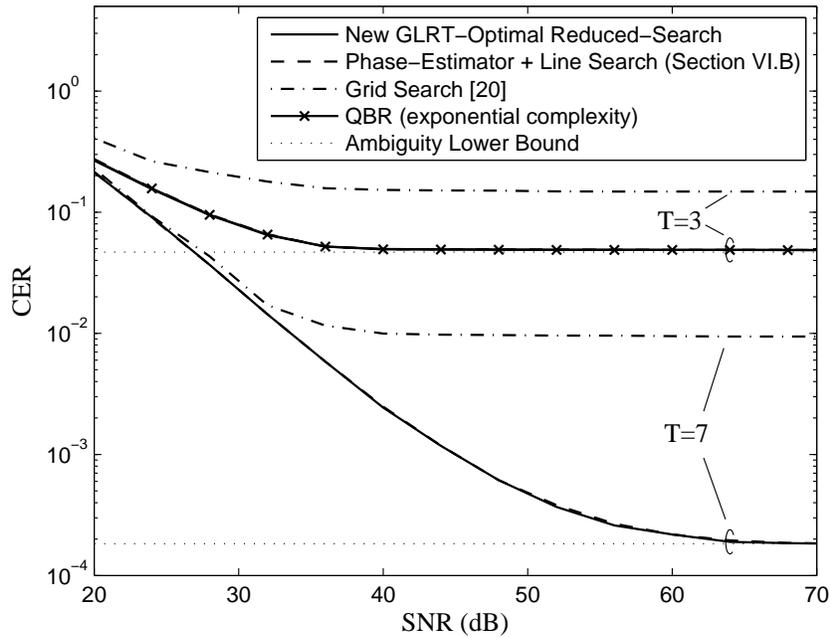}
\end{center}
\caption{\small Plot of Codeword Error Rate (CER) as a function of SNR for a 16-ary square QAM system.}
\label{fig:qam_bler}
\end{figure}

Figure \ref{fig:qam_bler} presents the CER as a function of SNR, for 16-QAM transmission. Results are shown for the GLRT-optimal QAM algorithm given in Section \ref{sec:QAM_algorithm} and the suboptimal algorithm given in Section \ref{sec:SubQAMalgorithm}. We also compare with the grid-based algorithm, where best performance for a fixed number of codeword estimates was obtained using $L=4$ channel phase estimates, which we also use for QBR.

For both the PAM and QAM cases we see that the suboptimal line-search algorithms provide negligible performance loss compared to the GLRT-optimal algorithm. For the case of $T=3$, where QBR is computationally possible, there is a noticeable performance loss. As discussed in Section \ref{sec:Detection}, divisor ambiguities result in a lower bound on the CER. Expressions for these lower bounds were provided in \cite{RyanIT06} and are also shown in the figure. Clearly, for high SNR, both of our GLRT-optimal algorithms and both suboptimal algorithms  detection achieve these bounds for both PAM and QAM. As noted in \cite{Warrier02}, there is an inherent suboptimality introduced by quantizing the unbounded channel attenuation by employing a grid-search approach, and hence the performance is clearly inferior. Also, although QBR achieves near-optimal performance for $T=3$, since the complexity of QBR increases exponentially with $T$ is not possible to produce curves for $T=7$.

In Table \ref{tab:NC} we present the relative computational complexities of the algorithms for the simulations in terms of the average number of codewords examined. The numbers in brackets indicate the number of codewords examined by the search if the restrictions on the search region provided by Theorems \ref{th:PAM} and \ref{th:QAM} are not applied (and are therefore slightly greater than the worst case values given in (\ref{eq:NCQAM}) and (\ref{eq:NCPAM})). We see that the suboptimal phase-estimator plus line-search approaches examine far fewer codewords yet obtains near-optimal performance, and that the complexity of QBR quickly becomes infeasible with increasing $T$.

\begin{table}[ht]
\begin{center}
\begin{tabular}{|ll|rr|rr|r|r|}
\hline
&& \multicolumn{2}{|c|}{GLRT-Optimal}
& \multicolumn{2}{|c|}{Phase Estimator}
& \multicolumn{1}{|c|}{QBR}
& \multicolumn{1}{|c|}{Grid}\\
&& \multicolumn{2}{|c|}{Reduced Search}
& \multicolumn{2}{|c|}{+ Line Search}
& \multicolumn{1}{|c|}{}
& \multicolumn{1}{|c|}{Search}  \\
\hline
8-PAM &T = 3     & 132.3     &(173)  & ~7.3          & (10)  &   128     & 174  \\
\hline
8-PAM &T = 7     & 772.6     &(1093) &  16.4         & (22)  &  32768    & 1094 \\
\hline
16-QAM &T = 3    &  ~~~52.6  &(87)   &  ~~~~22.9    &(28)   &  108      & 88\\
\hline
16-QAM &T = 7    & ~~~311.8  &(439)  &  ~~~~52.9    &(60)   &  8748     & 440\\
\hline
\end{tabular}
\caption{\label{tab:NC} Number of codewords examined for noncoherent PAM and QAM detection }
\end{center}
\end{table}

\section{Reduced Ambiguity Transmission} \label{sec:RA}

In this section we extend our new noncoherent detection algorithm to pilot assisted transmission (PAT) systems \cite{Tong04}. Unlike, standard PAT we propose to use the pilot symbol for noncoherent ambiguity resolution, rather than simply for channel estimation. We propose to replace the pilot symbol of PAT with a symbol generated in the following way. Two bits are allocated for resolving the $\pi/2$ phase ambiguity of square QAM, and the remaining bits in the symbol are allocated to parity, remove divisor ambiguities and improve error performance. Therefore, this scheme has the same data rate as PAT and can be compared directly.

With parity check bits in the codeword, we can now even further reduce the search space of our reduced search GLRT lattice-decoding algorithm by only considering codewords which satisfy a parity check. This significantly reduces the ambiguity problem. We will denote this parity-aided transmission scheme as reduced ambiguity (RA) transmission.

An arbitrarily chosen parity check scheme might reduce the number of divisor ambiguities, however since the metric (\ref{eq:x_ML}) has a geometric interpretation it may be possible to design other parity-check schemes which both resolve ambiguities and optimize performance by providing a minimum angular separation between codewords. The resolution of ambiguities can be achieved, at least for 16-QAM, by using the following parity-check scheme. Two parity bits $p_1,p_2$ are calculated from the data bits $\{ d_1,d_2,\ldots,d_{2(T-1)} \}$ as follows,
\begin{align}
p_1 &\equiv 1 + \sum_{t=1}^{4(T-1)}{d_{t}}  \label{eq:parity1}\\
p_2 &\equiv 1 + \sum_{t=1}^{2(T-1)}{d_{2t}} \label{eq:parity2}
\end{align}
where $\equiv$ denotes equality in $GF(2)$. They are then mapped to the upper right-hand quadrant of the QAM constellation of the first (pilot) symbol in the codeword as follows: $(00) \rightarrowtail 1+j$, $(01) \rightarrowtail 1+3j$, $(11) \rightarrowtail 3+3j$ and $(10) \rightarrowtail 3+j$. Effectively this means the first two bits of the first symbol of each codeword is chosen such that $\xub_1,\xub_2 > 0$, which removes the $\pi/2$ phase ambiguity, and the other two bits are parity bits, which in this case can be shown to completely remove the divisor ambiguities (see Appendix \ref{app:resolve}).

Figure \ref{fig:pvp_ber} presents the bit error rate (BER) as a
function of SNR for detection of 16-QAM transmitted over a block
independent phase-noncoherent AWGN channel. Again we have assumed
that the phase ambiguities have been removed within each codeword.
Results are shown for three codeword lengths $T=3,5,7$.  The figure
shows curves for our new RA reduced-search GLRT-optimal algorithm, and
compares them to standard PAT. Both schemes use a single pilot
symbol per codeword; which for the RA scheme is generated as
described above, and for PAT it is a symbol which has energy equal
to the average energy per symbol. For PAT, the GLRT estimate of the
channel (based on the pilot symbol) is used to perform GLRT-optimal data
detection, while for RA lattice decoding we use our reduced search
GLRT-optimal algorithm. Note that for PAT, the BER is independent of the
codeword length $T$ since it is a symbol-by-symbol detection scheme,
whereas for RA lattice decoding the BER decreases as $T$ increases
since it is a sequence detection scheme. Clearly our scheme
outperforms PAT increasingly with $T$.

\begin{figure}
\begin{center}
\includegraphics[width=0.7\linewidth]{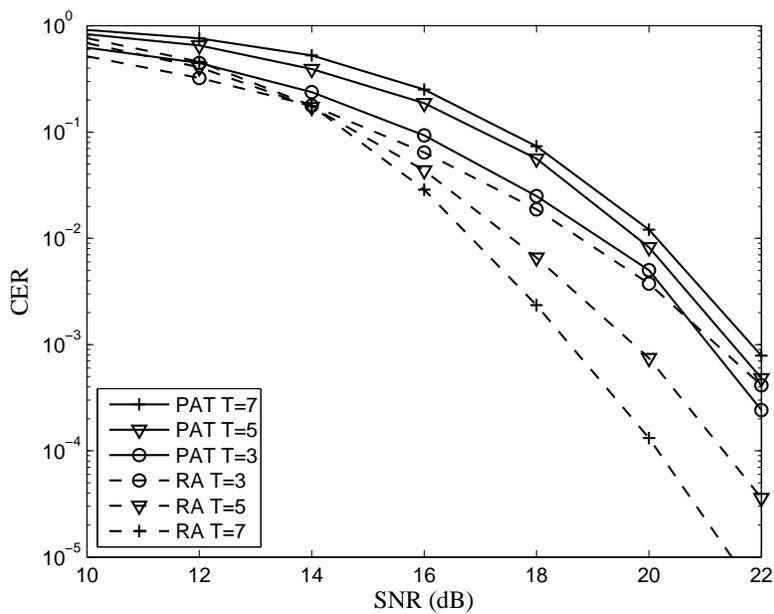}
\end{center}
\caption{\small Comparison of Codeword Error Rate (CER) as a function of SNR for 16-QAM for PAT versus RA transmission.}
\label{fig:pvp_bler}
\end{figure}

Figure \ref{fig:pvp_bler} shows the CER for the scenario of Figure
\ref{fig:pvp_ber}.  This serves to highlight even further the
benefit from our lattice (sequence) decoding approach compared with
PAT. For PAT, since bit errors occur independently on a
symbol-by-symbol basis, the CER increases with $T$. However, for RA
lattice decoding the CER decreases.  Also the figure highlights the
advantage of using pilot symbols, compared with fully noncoherent
transmission, by observing that the SNR range is significantly lower
than for Figures \ref{fig:pam_bler} and \ref{fig:qam_bler}.

\section{Conclusion}
In this paper we developed polynomial-time lattice-decoding algorithms for noncoherent block detection of PAM and QAM. Faster suboptimal algorithms for QAM were also presented which have excellent agreement with the optimal algorithms. A reduced ambiguity transmission scheme was introduced which was shown to outperform pilot assisted transmission over the phase noncoherent channel.

\begin{appendix}

\subsection{Strictly Optimal Calculation of Interior Points} \label{app:exactoptimality}
For each non-empty region $P(\xhubf)$, there exists a vertex $\nubf \in V(\xhubf)$ and small scalars $\nu^+,\nu^- > 0$, such that either $\nubf^+ \defas \nubf + [\;\nu^+ \;0\;]'$ or $\nubf^- \defas \nubf + [\;\nu^- \;0\;]'$ is in the interior of $P(\xhubf)$.

Suppose the first case is true. Now, the line $\nubf + \gamma [\; 1 \;0 \;]'$intersects an edge of the boundary of $P(\xubbf)$, and we will call this intersection point $\mubf$. We propose to choose $\nu^+=\gamma>0$ so that $\nubf^+$ is the midpoint of $\nubf$ and $\mubf$. Defining $\uub_t$ as the $t$th element of $\uubbf = \Yubbf \, \nubf$, where $\Yubbf$ is defined by the original received vector $\ybf$, we can calculate $\nu^+$ as follows,
\begin{equation*}
\nu^+ =
\begin{cases}
\substack{\min \\ t} \frac{ 2 \lceil \tfrac{\uub_t}{2} \rceil  - \uub_t }{ 2 \yub_t} &
\yub_t < 0, \\
\substack{\min \\ t} -\frac{ 2 \lfloor \tfrac{\uub_t}{2} \rfloor  - \uub_t }{ 2 \yub_t} &
\yub_t > 0.
\end{cases}
\end{equation*}
Note that almost surely $\yub_t \neq 0$. Similarly, using the line $\nubf - \gamma [\; 1 \;0 \;]'$ we calculate $\nu^- = \gamma > 0$ as
\begin{equation*}
\nu^- =
\begin{cases}
\substack{\min \\ t} -\frac{ 2 \lfloor \tfrac{\uub_t}{2} \rfloor  - \uub_t }{ 2 \yub_t} &
\yub_t < 0, \\
\substack{\min \\ t} \frac{ 2 \lceil \tfrac{\uub_t}{2} \rceil  - \uub_t }{ 2 \yub_t} &
\yub_t > 0.
\end{cases}
\end{equation*}
This process will in general always calculate a point in each non-empty $P(\xhubf)$. However, to avoid calculation problems we first rotate $\nu$ by $y_m/\abs{y_m}$, so that the vectors $[\;\nu^+ \;0\;]'$ and $[\;\nu^- \;0\;]'$ are not parallel to any of the edges of $P(\xhubf)$ (\eg those that are part of $\Scal$). This rotation is later reversed, so that the points calculated are in the original coordinates.

\subsection{Removal of Ambiguities in 16-ary QAM} \label{app:resolve}
In this section we show that the proposed RA pilot symbol approach (using parity checks, as discussed in Section \ref{sec:RA}) totally removes both the phase \emph{and} divisor ambiguities otherwise inherent in a noncoherent detection system (as discussed in Section \ref{sec:SigModel}).
\\
We start by recalling that the proposed parity scheme involves calculating the parity bits from the data bits $d_t$ for $t = 1,\ldots,4T$ as follows.
\begin{equation} \label{eq:parity}
p_1 \equiv 1 + \sum_{\ell=1}^{4(T-1)}{d_t} \;\;\;\;\;\;\;\; p_2 \equiv 1 + \sum_{\ell=1}^{2(T-1)}{d_{2t}}
\end{equation}
where $\equiv$ denotes equality in $GF(2)$.

The data and parity bits are then mapped to the symbols as shown in Table \ref{tab:bitmappings}, where we recall from the definition in (\ref{eq:c2r}) that $\xub_{2t-1} = \Real{x_t}$ and $\xub_{2t} = \Imag{x_t}$.
\begin{table}[ht]
\begin{center}
\begin{tabular}{r|r}
  $d_{2t-1} d_{2t}$ & $\xub_t$ \\
  \hline
  00 & $-3$ \\
  01 & $-1$ \\
  11 & $1$ \\
  10 & $3$ \\
\end{tabular}
\hspace{1.5cm}
\begin{tabular}{r|l}
  $p_1 p_2$ & $x_1$ \text{(pilot symbol)} \\
  \hline
  00 & $1+i$ \\
  01 & $1+3i$ \\
  11 & $3+3i$ \\
  10 & $3+i$ \\
\end{tabular}
\caption{\label{tab:bitmappings} Mapping of data and parity bits.}
\end{center}
\end{table}

Since $x_1$ is constrained to have positive real and imaginary components, the phase ambiguity has been removed. It remains to show that all divisor ambiguities have also been removed.

To do this, we first define the associates of a Gaussian integer $g$ to be the elements of the set $A(g) = \{g, gi, -g, -gi \}$. We also denote $A(g)^T$ to be a codeword of length $T$ composed of only elements of $A(g)$. For 16-QAM, it can be easily shown that a necessary condition for a divisor ambiguity to exist is that there exists codewords $\xbf^{(1)} \in A(g_1)^T$, $\xbf^{(2)} \in A(g_2)^T$ for some $g_1,g_2 \in \Xcal \defas \{ 1+i,3+3i,3+i,1+3i \}$ such that $g_1 \neq g_2$.

For a codeword $\xbf$ and some $g \in \Xcal$, we define $N_1,N_2,N_3$ and $N_4$ as the number of occurrences in a codeword of each of the four possible rotations of $g$ in the codeword, that is $g, gi, -g$ and $-gi$ respectively.
Noting that the phase ambiguity has been removed (since $x_1$ is constrained to have positive real and imaginary components), a sufficient condition for two codewords to be unambiguous is that there exists some $t$, such that the $t$th symbols from the two codewords are in different quadrants of the complex plane. It follows then, that a sufficient condition for two codewords to be unambiguous is that they do not have the same values of $N_1$ to $N_4$.

We now use this property on $N_1$ to $N_4$ to show that for arbitrary $T$, it is not possible for two ambiguous codewords $\xbf^{(1)} \in A(g_1)^T$, $\xbf^{(2)} \in A(g_2)^T$, to satisfy the parity check (\ref{eq:parity}) for any $g_1,g_2 \in \Xcal$ such that $g_1 \neq g_2$.

We consider each $g \in \Xcal$ (where we have previously defined $\Xcal = \{ 1+i,3+3i,3+i,1+3i \}$) in turn, showing that all codewords $\xbf \in A(g)^T$ that satisfy the parity check, are distinguishable in phase from all parity-satisfying codewords considered up to that point. For 16-QAM this process involves considering the four Gaussian integers $1+i$, $3+3i$, $3+i$ and $1+3i$ in turn, as detailed in the following four cases.

Define $\xbf^D$ to be the data codeword component of $\xbf$, i.e. $\xbf =  [x_2 \ldots x_T]'$.

\begin{itemize}
\item \textbf{Case $\xbf^D \in A(1+i)^{T-1}$}: \\
In this case, we show that there does not exist any $\xbf \in A(1+i)^T$ that satisfies the parity check.
Using Table \ref{tab:bitmappings}, the bits $(d_{4\ell-3} \ldots d_{4\ell})$ are mapped to the symbol $x_\ell = \xub_{2\ell-1} + i\xub_{2\ell} \in \xbf^D$ in the following way: $(1111) \rightarrowtail 1+i, (0111) \rightarrowtail -1+i, (0101) \rightarrowtail -1-i$ and $(1101) \rightarrowtail 1-i$. Clearly from (\ref{eq:parity2}), $p_2 \equiv 1$, and therefore the pilot symbol $x_1$ will be either $1+3i$ or $3+3i$. It follows that $\xbf \notin A(1+i)^T$.
\item \textbf{Case $\xbf^D \in A(3+3i)^{T-1}$}: \\
In this case, we show the conditions under which a codeword $\xbf \in A(3+3i)^T$ satisfies the parity check.
The associated bit mappings are $(1010) \rightarrowtail 3+3i, (0010) \rightarrowtail -3+3i, (0000) \rightarrowtail -3-3i$ and $(1000) \rightarrowtail 3-3i$. Clearly, $p_1 \equiv 1 + N_2 + N_4$ and $p_2 \equiv 1$.
Therefore,
\begin{equation*}
x_1 =
\Bigg \lbrace
\begin{matrix}
1+3i, &
\text{if } (p_1p_2) = (01) \text{\emph{~~i.e.} if~~} N_2 \not \equiv N_4, \\
3+3i, &
\text{if } (p_1p_2) = (11) \text{\emph{~~i.e.} if~~} N_2 \equiv N_4. \\
\end{matrix}
\end{equation*}
Furthermore it follows that $\xbf \in A(3+3i)^T$ only if $N_2 \equiv N_4$.

\item \textbf{Case $\xbf^D \in A(3+i)^{T-1}$}: \\
In this case, we show the conditions under which a codeword $\xbf \in A(3+i)^T$ satisfies the parity check, and show that under these conditions there does not exist any ambiguous codeword from $A(3+3i)^T$, i.e. from the previous case. The bit mappings are $(1011) \rightarrowtail 3+i, (0110) \rightarrowtail -1+3i, (0001) \rightarrowtail -3-i$ and $(1100) \rightarrowtail 1-3i$. In this case, $p_1 \equiv 1 + N_1 + N_3$ and $p_2 \equiv 1+N_1+N_2+N_3+N_4 \equiv 1+T-1 \equiv T$. If $T$ is odd, then $p_2 \equiv 1$ and therefore $x_1 \in \{1+3i,3+3i\}$ and therefore $\xbf \notin A(3+i)^T$ . If $T$ is even then $p_2 \equiv 0$ and $p_1 \equiv 1 + N_1 + N_3 \equiv N_2 + N_4$.
Therefore
\begin{equation*}
x_1 =
\Bigg \lbrace
\begin{matrix}
3 + i &
\text{if } (p_1 p_2) = (10)  \text{\emph{~~i.e.} if~~} N_2 \not \equiv N_4, \\
1+i, &
\text{if } (p_1 p_2) = (00) \text{\emph{~~i.e.} if~~} N_2 \equiv N_4. \\
\end{matrix}
\end{equation*}
It follows that $\xbf \in A(3+i)^T$ only if $N_2 \not \equiv N_4$ and $T$ is even. Recall that in the previous case, valid parity satisifying codewords only occurred if $N_2 \equiv N_4$. Therefore an ambiguity will not occur between two codewords $\xbf \in A(3+i)^T$ and $\xbf^{(1)} \in A(3+3i)^T$ since they will be distinguishable in phase.

\item \textbf{Case $\xbf^D \in A(1+3i)^{T-1}$}:\\
In this case, we show the conditions under which a codeword $\xbf \in A(1+3i)^T$ satisfies the parity check, and show that under these conditions there does not exist any ambiguous codeword from either $A(3+3i)^T$ or $A(3+i)^T$, i.e. from the previous two cases. The bit mappings are $(1110) \rightarrowtail 1+3i, (0011) \rightarrowtail -3+i, (0100) \rightarrowtail -1-3i$ and $(1001) \rightarrowtail 3-i$. Here, $p_1 \equiv 1 + N_1 + N_3$ and $p_2 \equiv T$. If $T$ is even, then $p_2 \equiv 0$ and therefore $x_1 \in \{1+i,3+i\}$ and no ambiguity occurs. If $T$ is odd then $p_1 \equiv 1 + N_2 + N_4$ and $p_2 \equiv 1$. Therefore,
\begin{equation*}
x_1 =
\Bigg \lbrace
\begin{matrix}
1 + 3i &
\text{if }(p_1 p_2) = (01)  \text{\emph{~~i.e.} if~~}  N_2 \not \equiv N_4, \\
3 + 3i, &
\text{if }(p_1 p_2) = (11)  \text{\emph{~~i.e.} if~~}  N_2 \equiv N_4.
\end{matrix}
\end{equation*}
It follows that $\xbf \in A(1+3i)^T$ only if $N_2 \not \equiv N_4$ and $T$ is odd. Clearly, these conditions are different to those to the previous two cases and therefore no ambiguities exist.
\end{itemize}

\end{appendix}

\bibliographystyle{IEEEtran}
\bibliography{journals-abbrev,dans}

\begin{thebibliography}{10}
\providecommand{\url}[1]{#1}
\csname url@rmstyle\endcsname
\providecommand{\newblock}{\relax}
\providecommand{\bibinfo}[2]{#2}
\providecommand\BIBentrySTDinterwordspacing{\spaceskip=0pt\relax}
\providecommand\BIBentryALTinterwordstretchfactor{4}
\providecommand\BIBentryALTinterwordspacing{\spaceskip=\fontdimen2\font plus
\BIBentryALTinterwordstretchfactor\fontdimen3\font minus
  \fontdimen4\font\relax}
\providecommand\BIBforeignlanguage[2]{{%
\expandafter\ifx\csname l@#1\endcsname\relax
\typeout{** WARNING: IEEEtran.bst: No hyphenation pattern has been}%
\typeout{** loaded for the language `#1'. Using the pattern for}%
\typeout{** the default language instead.}%
\else
\language=\csname l@#1\endcsname
\fi
#2}}

\bibitem{Marzetta99}
T.~L. Marzetta and B.~M. Hochwald, ``Capacity of a mobile multiple-antenna
  communication link in {R}ayleigh flat fading,'' \emph{IEEE Trans. Inform.
  Theory}, vol.~45, no.~1, pp. 139--157, Jan. 1999.

\bibitem{Zheng02}
L.~Zheng and D.~N.~C. Tse, ``Communication on the {G}rassmann manifold: A
  geometric approach to the noncoherent multiple-antenna channel,'' \emph{IEEE
  Trans. Inform. Theory}, vol.~48, pp. 359--383, Feb. 2002.

\bibitem{Hassibi03}
B.~Hassibi and B.~M. Hochwald, ``How much training is needed in
  multiple-antenna wireless links?'' \emph{IEEE Trans. Inform. Theory},
  vol.~49, no.~4, pp. 951--963, Apr. 2003.

\bibitem{Chen03}
R.-R. Chen, R.~Koetter, D.~Agrawal, and U.~Madhow, ``Joint demodulation and
  decoding for the noncoherent block fading channel: A practical framework for
  approaching {S}hannon capacity,'' \emph{IEEE Trans. Commun.}, vol.~51, pp.
  1676--1689, Oct. 2003.

\bibitem{Cavers91}
J.~K. Cavers, ``An analysis of pilot symbol assisted modulation for {R}ayleigh
  fading channels,'' \emph{IEEE Trans. Veh. Technol.}, vol.~40, no.~4, pp.
  686--693, 1991.

\bibitem{Tong04}
L.~Tong, B.~M. Sadler, and M.~Dong, ``Pilot-assisted wireless transmissions,''
  \emph{IEEE Signal Processing Mag.}, vol.~21, pp. 12--25, Nov. 2004.

\bibitem{Raheli95}
R.~Raheli, A.~Polydoros, and C.-K. Tzou, ``Per-survivor processing: A general
  approach to {MLSE} in uncertain environments,'' \emph{IEEE Trans. Commun.},
  vol.~43, no. 2/3/4, pp. 354--364, Feb 1995.

\bibitem{Davis01b}
L.~M. Davis, I.~B. Collings, and P.~Hoeher, ``Joint {MAP} equalization and
  channel estimation for frequency-selective and frequency-flat fast-fading
  channels,'' \emph{IEEE Trans. Commun.}, vol.~49, no.~12, pp. 2106--2114,
  December 2001.

\bibitem{Georghiades97}
C.~N. Georghiades, ``Blind carrier phase acquisition for {QAM}
  constellations,'' \emph{IEEE Trans. Commun.}, vol.~45, no.~11, pp.
  1477--1486, Nov. 1997.

\bibitem{Weber78}
W.~E. Weber, ``Differential encoding for multiple amplitude and phase shift
  keying systems,'' \emph{IEEE Trans. Commun.}, vol.~26, pp. 385--391, Mar.
  1978.

\bibitem{Warrier02}
D.~Warrier and U.~Madhow, ``Spectrally efficient noncoherent communication,''
  \emph{IEEE Trans. Inform. Theory}, vol.~48, pp. 652--668, Mar. 2002.

\bibitem{Lampe05}
L.~Lampe, R.~Schober, V.~Pauli, and C.~Windpassinger, ``Multiple-symbol
  differential sphere decoding,'' \emph{IEEE Trans. Commun.}, vol.~53, pp.
  1981--1985, Dec. 2005.

\bibitem{ClarksonK01}
K.~L. Clarkson, W.~Sweldens, and A.~Zheng, ``Fast multiple-antenna differential
  decoding,'' \emph{IEEE Trans. Commun.}, vol.~49, no.~2, pp. 253--261, Feb.
  2001.

\bibitem{CongLing05}
C.~Ling, W.~H. Mow, K.~H. Li, and A.~C. Kot, ``Multiple-antenna differential
  lattice decoding,'' \emph{IEEE J. Sel. Areas Commun.}, vol.~23, no.~9, pp.
  1821--1829, Sept. 2005.

\bibitem{RyanISIT05}
D.~J. Ryan, I.~V.~L. Clarkson, and I.~B. Collings, ``Detection error
  probabilities in noncoherent channels,'' in \emph{Proc. IEEE Int. Symp. on
  Inform. Theory (ISIT)}, Adelaide, Australia, Sept. 2005, pp. 617--621.

\bibitem{PauliISIT06}
V.~Pauli and L.~Lampe, ``On the complexity of sphere decoding for {MSDD},'' in
  \emph{Proc. IEEE Int. Symp. on Inform. Theory (ISIT)}, Seattle, WA, July
  2006, pp. 932--936.

\bibitem{Agrell2002}
E.~Agrell, T.~Eriksson, A.~Vardy, and K.~Zeger, ``Closest point search in
  lattices,'' \emph{IEEE Trans. Inform. Theory}, vol.~48, no.~8, pp.
  2201--2214, Aug. 2002.

\bibitem{Mackenthun}
K.~M. {Mackenthun Jr.}, ``A fast algorithm for multiple-symbol differential
  detection of {MPSK},'' \emph{IEEE Trans. Commun.}, vol.~42, pp. 1471--1474,
  Feb./Mar./Apr. 1994.

\bibitem{Sweldens01}
W.~Sweldens, ``Fast block noncoherent decoding,'' \emph{IEEE Comms. Letters},
  vol.~5, no.~4, pp. 132--134, Apr. 2001.

\bibitem{Motedayen03}
I.~Motedayen-Aval and A.~Anastasopoulos, ``Polynomial-complexity noncoherent
  symbol-by-symbol detection with application to adaptive iterative decoding of
  turbo-like codes,'' \emph{IEEE Trans. Commun.}, pp. 197--207, Feb. 2003.

\bibitem{MotedayenICC02}
------, ``Polynomial-complexity, adaptive symbol-by-symbol soft-decision
  algorithms with application to non-coherent detection of {LDPCC},'' in
  \emph{Proc. IEEE Int. Conf. on Communications (ICC)}, New York, USA, Apr.
  2002, pp. 1677--1681.

\bibitem{MotedayenICC03}
------, ``Polynomial complexity {ML} sequence and symbol-by-symbol detection in
  fading channels,'' in \emph{Proc. IEEE Int. Conf. on Communications (ICC)},
  Anchorage, Alaska, May 2003, pp. 2718--2722.

\bibitem{VanTrees}
H.~L. {Van Trees}, \emph{Detection, Estimation, and Modulation Theory: Part
  I}.\hskip 1em plus 0.5em minus 0.4em\relax John Wiley \& Sons, 1968.

\bibitem{Wong67}
Y.-C. Wong, ``Differential geometry of {G}rassmann manifolds,'' \emph{Proc.
  Nat. Acad. Sci. USA}, vol.~47, pp. 589--594, 1967.

\bibitem{RyanIT06}
D.~J. Ryan, I.~V.~L. Clarkson, and I.~B. Collings, ``Blind detection of {PAM}
  and {QAM} in fading channels,'' \emph{IEEE Trans. Inform. Theory}, vol.~52,
  pp. 1197--1206, Mar. 2006.

\bibitem{Cormen01}
T.~H. Cormen, C.~E. Leiserson, R.~L. Rivest, and C.~Stein, \emph{Introduction
  to Algorithms}, 2nd~ed.\hskip 1em plus 0.5em minus 0.4em\relax Cambridge, MA,
  USA: McGraw-Hill Higher Education, 2001.

\bibitem{Moen94}
M.~Moeneclaey and G.~{de Jonghe}, ``{ML}-oriented {NDA} carrier synchronization
  for general rotationally symmetric signal constellations,'' \emph{IEEE Trans.
  Commun.}, pp. 2531--2533, Aug. 1994.

\end{thebibliography}

\renewcommand{\baselinestretch}{1}
\setcounter{a}{0}
\begin{table}[ht]
\begin{center}
\begin{tabular}{|l|l|}
  \hline
  \algnum   \unbf{begin} \\
  \algnum   ~~$\sbf  := \sgn{\ybf}$;
  \hfill// Store sign of each $y_t$\\
  \algnum   ~~$\ybf  := \sbf \circ \ybf$;
  \hfill// Make each $y_t$ positive \\
  \algnum   ~~$\Bmax := M+T-2$;  \\
  \algnum   ~~$m  := \arg \max_t \cubr{y_t}$; \\
  \algnum   ~~$\lambdamax := (M + 2T - 2)/\abs{y_m}$;
  \hfill// Search region: $0 < \lambda < \lambdamax$\\
  \algnum   ~~$\Vcal_0 := \emptyset$;
  \hfill// Calculate and store $P(\xubbf)$ boundary points \\
  \algnum   ~~\unbf{for} $t := 1$ \unbf{to} $T$ \unbf{do} \\
  \algnum   ~~~~\unbf{for} \unbf{all} $b \in \cubr{2,4,\ldots,M-2}$ \unbf{do}\\
  \algnum   ~~~~~~$\nu := b/y_t$;\\
  \algnum   ~~~~~~\unbf{if} $\nu < \lambdamax$; \\
  \algnum   ~~~~~~~~$\Vcal_0 := \cubr{\Vcal,(\nu,t)}$;\\
  \algnum   ~~~~~~\unbf{else} \unbf{break}; \\
  \algnum   ~~~~\unbf{end} \unbf{for} \unbf{all};\\
  \algnum   ~~\unbf{end} \unbf{for;} \\
  \algnum   ~~$\Vcal := \text{sort}(\Vcal_0)$; \;\;
  \hfill// Sort $\Vcal_0$ in ascending order of $\nu$\\
  \algnum   ~~$\Vcal := \cubr{\Vcal,(\lambdamax,0)}$; \\
  \algnum   ~~$\xhbf := [\; 1\;1\;\ldots\;1\;]'$;
  \hfill// Initialize data estimate\\
  \algnum   ~~$\alpha := \xhbf' \ybf$;
  \hfill// Initialize likelihood terms\\
  \algnum   ~~$\beta  := \magn{\xhbf}^2 $; \\
  \algnum   ~~$L  := \alpha^2 / \beta $; \\
  \algnum   ~~$\lambda := \Vcal(1,1)/2$; \\
  \algnum   ~~\unbf{for} $k := 1$ \unbf{to} $\abs{\Vcal}-1$ \unbf{do}
  \hfill// Iteratively examine likelihoods \\
  \algnum   ~~~~$t  := \Vcal(k,2) $; \\
  \algnum   ~~~~$\alpha := \alpha + 2y_{t}$;
  \hfill// Update likelihood terms\\
  \algnum   ~~~~$\beta := \beta + 4\xh_{t} + 4$; \\
  \algnum   ~~~~$\xh_{t}  := \xh_{t} + 2 $;
  \hfill// Update $\xbf$\\
  \algnum   ~~~~\unbf{if} $\alpha^2 / \beta > L$
  \hfill// If better $\xbf$ found\\
  \algnum   ~~~~~~$L := \alpha^2 / \beta$;
  \hfill// Update likelihood \\
  \algnum   ~~~~~~$\lambda := (\Vcal(k,1)+\Vcal(k+1,1))/2$;
  \hfill// Store point in $P(\xhbf)$ \\
  \algnum   ~~~~\unbf{end} \unbf{if}; \\
  \algnum   ~~\unbf{end} \unbf{for}; \\
  \algnum   \unbf{return} $\xhbf^{\text{opt}} := \sbf \circ NN (\lambda \ybf )$;  \\
  \hline
\end{tabular}
\caption{$M$-ary real-PAM noncoherent lattice decoding algorithm}
\label{tab:algorithm}
\end{center}
\vspace{-0.5cm}
\end{table}
\renewcommand{\baselinestretch}{1.75}

\newpage
\renewcommand{\baselinestretch}{1}
\setcounter{a}{0}
\begin{table}
\begin{center}
\begin{tabular}{|l|l|}
  \hline
  \algnum  \unbf{begin}\\
  \algnum   ~~$m  := \arg \max_t \cubr{\abs{y_t}}$; \\
  \algnum   ~~$y  := (y_m^*/\abs{y_m}) \ybf$;
  \hfill // Rotate $\ybf$ so that $y_m$ is purely real  \\
  \algnum   ~~$\Bmax := M+2T-2$; \\
  \algnum   ~~$\lambdamax := \Bmax/\abs{y_m}$;
  \hfill // Search boundary (Thm. \ref{th:QAM}).  \\
  \algnum   ~~$\xbest := NN((\epsilon+i\epsilon)\ybf )$;
  \hfill // Codeword near origin\\
  \algnum   ~~$\Lmax := L(\xbest)$;
  \hfill // L'hood $L(\xbf) \defas \abs{\xbf^\dag \ybf}^2/\magn{\xbf}^2$\\
  \algnum   ~~$\Bcal := \cubr{2,4,\ldots,M-2}$;
  \hfill // Positive NN boundaries\\
  \algnon   ~~// Calculate only intersection points (\ie  vertices) in  \\
  \algnon   ~~// first quadrant using (\ref{eq:Vcal_solve}) by reducing number of\\
  \algnon   ~~// NN boundaries $\Bcal_1,\Bcal_2$ and then rotating.\\
  \algnum   ~~\unbf{for} $t := 1$ \unbf{to} $2T-1$ \unbf{do}\\
  \algnum   ~~~~$\Bcal_1 := \Bcal$; \\
  \algnum   ~~~~\unbf{if} $t \in \cubr{2m - 1,2m}$ \unbf{then} $\Bcal_1 := \cubr{\Bcal_1,\Bmax}$; \\
  \algnum   ~~~~\unbf{for} $t' := t+1$ \unbf{to} $2T$ \unbf{do}\\
  \algnum   ~~~~~~$\Bcal_2 := \Bcal$; \\
  \algnum   ~~~~~~\unbf{if} $t' \in \cubr{2m - 1,2m}$ \unbf{then} $\Bcal_2 := \cubr{\Bcal_2,\Bmax}$; \\
  \algnum   ~~~~~~$\Sbf := \sqbr{ \begin{matrix} \Yub_{t,1} & \Yub_{t,2} \\ \Yub_{t',1} & \Yub_{t',2} \end{matrix} }^{-1}$;
  \hfill // Matrix in (\ref{eq:Vcal_solve})\\
  \algnum   ~~~~~~\unbf{for} \unbf{all} $b_1 \in \Bcal_1 $ \\
  \algnum   ~~~~~~~~\unbf{for} \unbf{all} $b_2 \in \Bcal_2$ \\
  \algnon   ~~~~~~~~~~// Calculate intersection point; \\
  \algnum   ~~~~~~~~~~$\nu$ :=  \text{Real-To-Complex}$( \Sbf [\; b_1 \; b_2 \;]' )$;\\
  \algnum   ~~~~~~~~~~$\nu$ :=  \text{Rotate-To-First-Quadrant}$(\nu)$;\\
  \algnum   ~~~~~~~~~~\unbf{for} \unbf{all} $s \in \cubr{-1,1}$ \\
  \algnum   ~~~~~~~~~~~~$\lambda :=  \nu + s(\epsilon+i\epsilon)$;
  \hfill // Point in some partition\\
  \algnon   ~~~~~~~~~~~~// Check that $\lambda$ is in reduced search region \\
  \algnum   ~~~~~~~~~~~~\unbf{if} $0 < \Real{\lambda} < \lambdamax$ \unbf{and} \\
  \algnon   ~~~~~~~~~~~~~~ $0 \leq  \Imag{\lambda} < \lambdamax$ \unbf{then} \\
  \algnum   ~~~~~~~~~~~~~~$\xhbf := NN(\lambda \ybf)$;
  \hfill // Calculate NN\\
  \algnum   ~~~~~~~~~~~~~~\unbf{if} $L(\xhbf) > \Lmax$
  \hfill // If better $\xbf$ found\\
  \algnum   ~~~~~~~~~~~~~~~~$\xbest := \xhbf$;
  \hfill // Update codeword estimate  \\
  \algnum   ~~~~~~~~~~~~~~~~$\Lmax := L(\xhbf)$;
  \hfill // Update likelihood \\
  \algnum   ~~~~~~~~~~~~~~\unbf{end} \unbf{if}; \\
  \algnum   ~~~~~~~~~~~~\unbf{end} \unbf{if}; \\
  \algnum  ~~~~~~~~~~\unbf{end} \unbf{for} \unbf{all}; \\
  \algnum  ~~~~~~~~\unbf{end} \unbf{for} \unbf{all}; \\
  \algnum  ~~~~~~\unbf{end} \unbf{for} \unbf{all}; \\
  \algnum  ~~~\unbf{end} \unbf{for}; \\
  \algnum  ~~\unbf{end} \unbf{for}; \\
  \algnum  \unbf{return} $\xML := \xbest$; \\
  \hline
\end{tabular}
\caption{$M^2$-ary square QAM noncoherent lattice decoding algorithm\label{tab:qam_algorithm} }
\end{center}
\vspace{-1cm}
\end{table}
\renewcommand{\baselinestretch}{1.75}

\newpage
\setcounter{a}{0}
\renewcommand{\baselinestretch}{1}
\begin{table}
\begin{center}
\begin{tabular}{|l|l|}
  \hline
  \algnum   \unbf{begin} \\
  \algnum   ~~$L := 0$; \hfill // Initialize likelihood \\
  \algnum   ~~$\lambdamax_0 := (M+2T-2)/\max_t \abs{y_t}$; \\
  \algnum   ~~\unbf{for} $\ell=1$ \unbf{to} $L$ \unbf{then} \\
  \algnum   ~~~~// Search region: $0 < \lambda < \lambdamax$ (Theorem \ref{th:QAM})\\
  \algnum   ~~~~$\lambdamax := \lambdamax_0 / \min \cubr{ \abs{\cos \frac{{\ell} \pi}{2L}},\abs{\sin \frac{{\ell} \pi}{2L}}}$;\\
  \algnum   ~~~~$\Vcal_0 = \emptyset;$
  \hfill // Calculate and store $P(\xubbf)$ boundary points \\
  \algnum   ~~~~\unbf{for} $t=1$ \unbf{to} $2T$ \unbf{then} \\
  \algnum   ~~~~~~\unbf{for} \unbf{all} $b \in \cubr{2,4,\ldots,M-2}$ \unbf{then} \\
  \algnum   ~~~~~~~~$\nu := b / |\yub_t|;$ \\
  \algnum   ~~~~~~~~\unbf{if} $\nu < \lambdamax$; \\
  \algnum   ~~~~~~~~~~$\Vcal_0 := \cubr{\Vcal_0,(\nu,t)}$;\\
  \algnum   ~~~~~~~~\unbf{else} \unbf{break}; \\
  \algnum   ~~~~~~\unbf{end} \unbf{for} \unbf{all};\\
  \algnum   ~~~~\unbf{end} \unbf{for;} \\
  \algnum   ~~~~$\Vcal := \text{sort}(\Vcal_0)$; \;\;
  \hfill // Sort $\Vcal_0$ in ascending order of $\nu$\\
  \algnum   ~~~~$\Vcal := \cubr{\Vcal,(\lambdamax,0)}$; \\
  \algnum   ~~~~$\subbf := \sgn \{\yubbf \}$; \\
  \algnum   ~~~~$\xhubf := \subbf$; \hfill // Initialize data estimate\\
  \algnum   ~~~~$\alpha := \xhbf^\dag \ybf$;
  \hfill// Initialize likelihood terms\\
  \algnum   ~~~~$\beta  := \magn{\xhbf}^2 $; \\
  \algnum   ~~~~\unbf{if} $\alpha^2 / \beta > L$
  \hfill// If better $\xbf$ found\\
  \algnum   ~~~~~~$L := \alpha^2 / \beta$;
  \hfill// Update likelihood \\
  \algnum   ~~~~~~$\lambda := \Vcal(1)/2$; \\
  \algnum   ~~~~~~$\phi := {\ell} \pi/(2L)$; \\
  \algnum   ~~~~\unbf{end} \unbf{if}; \\
  \algnum   ~~~~\unbf{for} $k := 1$ \unbf{to} $\abs{\Vcal}-1$ \unbf{do}
    \hfill// Iteratively examine likelihoods \\
  \algnum   ~~~~~~$t := \Vcal(k,2)$; \\
  \algnum   ~~~~~~\unbf{if} $t'$ is odd \unbf{then} \\
  \algnum   ~~~~~~~~$\alpha := \alpha + 2 \sub_{t} y_{(t+1)/2}$; \\
  \algnum   ~~~~~~\unbf{else} \\
  \algnum   ~~~~~~~~$\alpha := \alpha - 2i \sub_{t} y_{t/2}$; \\
  \algnum   ~~~~~~\unbf{end} \unbf{if} \\
  \algnum   ~~~~~~$\beta := \beta + 4 \sub_{t} \xhub_{t}+4$; \\\emph{}
  \algnum   ~~~~~~$\xhub_{t}  := \xhub_{t} + 2\sub_{t}$; \\
  \algnum   ~~~~~~\unbf{if} $\abs{\alpha}^2 / \beta > L$
  \hfill// If better $\xbf$ found\\
  \algnum   ~~~~~~~~$L := \abs{\alpha}^2 / \beta$;
  \hfill// Update likelihood\\
  \algnum   ~~~~~~~~$\lambda := (\Vcal(k,1)+\Vcal(k+1,1))/2$; \hfill// Store point in $P(\xhubf)$\\
  \algnum   ~~~~~~~~$\phi := \ell \pi/(2L)$; \\
  \algnum   ~~~~~~\unbf{end} \unbf{if}; \\
  \algnum   ~~~~\unbf{end} \unbf{for}; \\
  \algnum   ~~~~$y := y e^{\tfrac{j\pi}{2L}}$;
  \hfill // Rotate $\ybf$ for next line search \\
  \algnum   ~~\unbf{end} \unbf{for}; \\
  \algnum   \unbf{return} $\xhbf^{\text{opt}} := NN (\lambda e^{j \phi} \ybf )$;\\
  \hline
\end{tabular}\caption{Suboptimal $M^2$-ary square QAM multiple line-search noncoherent detection algorithm\label{tab:QAMSub} }
\end{center}
\vspace{-1cm}
\end{table}
\renewcommand{\baselinestretch}{1.75}

\newpage

\begin{figure}
\begin{center}
\includegraphics[width=0.7\linewidth]{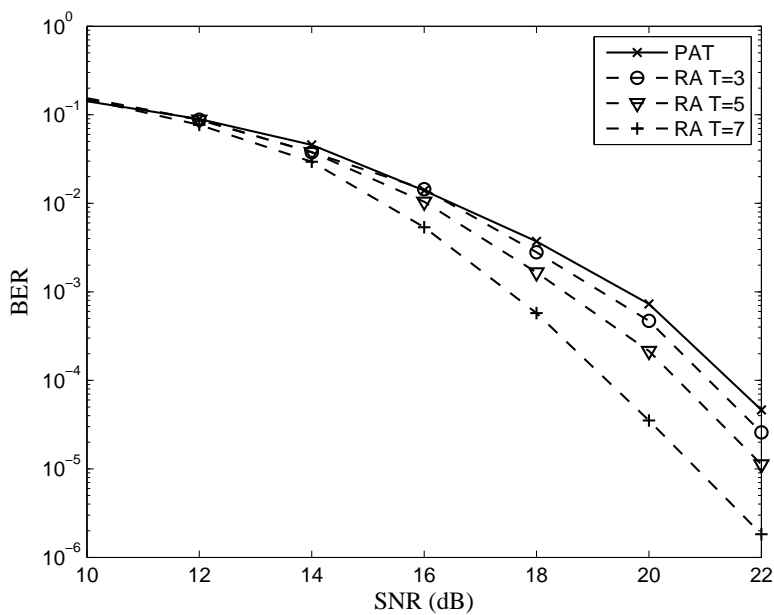}
\end{center}
\caption{\small Comparison of Bit Error Rate (BER) as a function of SNR for 16-QAM for PAT versus RA transmission.}
\label{fig:pvp_ber}
\end{figure}

\end{document}